\documentclass[12pt]{article}
\usepackage{savesym}
\usepackage{cite}
\usepackage{wasysym}
\usepackage{amscd}
\usepackage{graphicx}
\usepackage{pifont}
\usepackage{float}
\usepackage{subfig}
\usepackage[all]{xy}
\usepackage{multicol}
\usepackage{amsfonts}
\usepackage{picture}
\usepackage{amssymb}
\savesymbol{iint}
\savesymbol{iiint}
\usepackage{amsmath}
\restoresymbol{TXF}{iint}
\restoresymbol{TXF}{iiint}
\usepackage{color}
\usepackage{clay}
\usepackage{hyperref}
\usepackage{slashed}
\hypersetup{colorlinks=true}
\hypersetup{linkcolor=black}
\hypersetup{citecolor=black}
\hypersetup{urlcolor=black}
\usepackage{setspace}
\usepackage{multirow}
\usepackage{wrapfig}
\usepackage{verbatim}
\numberwithin{equation}{section}
\begin{document}
\date{May, 2013}

\institution{Fellows}{\centerline{${}^{1}$Society of Fellows, Harvard University, Cambridge, MA, USA}}
\institution{HarvardU}{\centerline{${}^{2}$Jefferson Physical Laboratory, Harvard University, Cambridge, MA, USA}}

\title{Complex Chern-Simons from M5-branes on the Squashed Three-Sphere}

\authors{Clay C\'{o}rdova,\worksat{\Fellows}\footnote{e-mail: {\tt cordova@physics.harvard.edu}}  and Daniel L. Jafferis \worksat{\HarvardU}\footnote{e-mail:{\tt jafferis@physics.harvard.edu}} }

\abstract{We derive an equivalence between the (2,0) superconformal M5-brane field theory dimensionally reduced on a squashed three-sphere, and Chern-Simons theory with complex gauge group.  In the reduction, the massless fermions obtain an action which is second order in derivatives and are reinterpreted as ghosts for gauge fixing the emergent non-compact gauge symmetry. A squashing parameter in the geometry controls the imaginary part of the complex Chern-Simons level.}

\maketitle

\setcounter{tocdepth}{2}
\tableofcontents
\section{Introduction}
\label{intro}

In this paper we show how to put the M5-brane theory on a squashed three-sphere while preserving four supercharges. The low energy limit of the resulting compactified theory in three dimensions is (equivalent to) complex Chern-Simons theory. The M5-brane field theory is labelled by an ADE Lie algebra $\mathfrak{g},$ and the appearance of $\mathfrak{g}$ type gauge fields is expected in compactification of the M5-brane theory.  In our construction, ``twisted'' scalars make up the remainder of $\mathfrak{g}_{\mathbb{C}}$. Fascinatingly, the gauginos end up with a second order action, and are reinterpreted as Faddeev-Popov ghosts of gauge fixing an emergent noncompact part of the gauge symmetry.

In the study of Chern-Simons theory it is often useful to add the Yang-Mills term as a regulator. This does not affect the IR limit, but renders the Euclidean path integral bounded, rather then merely oscillatory. However, in complex Chern-Simons \cite{Witten, Dimofte:2009yn, Witten:2010cx} this cannot be done, since the Yang-Mills term for a complex gauge group has indefinite sign making the action unbounded from below. In our setup, the system is well-defined before taking the $r \rightarrow 0$  compactification limit. The relation to complex Chern-Simons requires changing the contour of integration of one of the fields. Turned around, this gives the nonperturbative definition of the complex Chern-Simons path integral in our context. 

With this understanding of a non-perturbative completion, our final three-dimensional action may be written simply in terms of a $\mathfrak{g}_{\mathbb{C}}$ connection $\mathcal{A}=A+iX$, and takes the form    
 \begin{eqnarray}
S & = & \frac{q}{8\pi} \int \mathrm{Tr}\left({\cal A} \wedge d {\cal A} + \frac{2}{3} {\cal A} \wedge {\cal A} \wedge {\cal A}\right) + \frac{\tilde{q}}{8\pi} \int \mathrm{Tr}\left(\bar{\cal A} \wedge d\bar{\cal A} + \frac{2}{3} \bar{\cal A} \wedge \bar{\cal A} \wedge\bar{\cal A} \right) \label{SCS} \\  
 & = & \frac{k}{4\pi} \int \mathrm{Tr}\left(A\wedge dA +\frac{2}{3} A^3 - X \wedge d_A X \right) + \frac{u}{2\pi} \int\mathrm{Tr} \left( \frac{1}{3} X^3- X \wedge F_A  \right), \nonumber
 \end{eqnarray} 
 where $q = k + iu$ and $\tilde{q} = k - i u$ for $k$ and $u$ not necessarily real. There are complex gauge transformations, ${\cal A} \rightarrow {\cal A} + d_{\cal A} g$, for $g \in \frak{g}_{\mathbb{C}}$ which in terms of $A$ and $X$ look like an ordinary $\mathfrak{g}$ gauge transformation of $A$ under which $X$ is an adjoint, and an additional noncompact gauge redundancy acting as $A \rightarrow A - [X, h]$ and $X \rightarrow X +  d_A h$. 

Invariance under large gauge transformations requires that $k$ be an integer. However, $u$ is not subject to any quantization condition.  There are two branches of values of $u$ that respect unitarity: either $u$ is real or pure imaginary depending on whether $X$ is taken to be even or odd under parity \cite{Witten}. In our compactification we find that the levels take the values 
\begin{equation}
k=1, \hspace{.5in}u=\sqrt{1-\ell^{2}}, \label{leveldefintro}
\end{equation}
where in the above $\ell \in \mathbb{R}_{+}$ is a squashing parameter in the geometry.  Thus, depending on whether $\ell$ is smaller or larger than unity, one lands on one or the other unitary branch.

The existence of the $(2,0)$ superconformal field theory \cite{Witten:1995zh, Strominger:1995ac, Witten:1995em}  in six dimensions provides powerful unifying principles for lower dimensional supersymmetric quantum field theories, and our result has implications in that context \cite{Witten:1997sc, Gaiotto:2009we}.  A rich class of 3d ${\cal N}=2$ superconformal field theories arise from compactifying the 6d $(2,0)$ theory on three-manifolds \cite{Dimofte:2011ju, Cecotti:2011iy, Cordova:2012xk, Dimofte:2013iv}. One quarter of the supersymmetry is preserved by adjusting background R gauge fields on the three-manifold, $M_3$, such that there exist twisted covariantly constant spinors. After Kaluza-Klein reduction, the resulting theory in $\mathbb{R}^{2,1}$  typically flows to an interacting superconformal field theory, $T_\frak{g}(M_3)$ in the IR. This IR SCFT does not depend on the metric of the three-manifold.  For some theories in this class, there are proposals for 3d Lagrangian descriptions as abelian Chern-Simons-matter theories. 

In general, when coupling a quantum field theory to a curved background metric, one has the option of adding additional terms to the action that disappear in the flat space limit. By dimensional counting these must be background values for the coefficients of relevant operators. For example, one may have curvature couplings to mass terms of the form ${\cal R} \phi^2$.   If we want the coupling to preserve supersymmetry then the best way to organize additional terms in the curved background is to imagine weakly coupling the field theory to off-shell supergravity \cite{Festuccia:2011ws}. Then one looks for configurations of the supergravity fields, without imposing on them any equation of motion or on-shell condition, that are invariant under some rigid supersymmetry. The role of the supergravity fields is to keep track of the additional terms in the coupled quantum field theory that are needed to preserve supersymmetry in a given fixed background geometry. 

The procedure of twisting on $M_3$ to construct the theory $T_\frak{g}(M_3)$ is an example of preserving supersymmetry in curved backgrounds where the only supergravity fields with non-trivial profiles are the metric and R gauge field. There are obviously no covariantly constant spinors on a general three-manifold, however one may tune the $so(3) \subset so(5)_{R}$ part of the R gauge field so that the spin and R connections exactly cancel for $1/4$ of the spinors, leading to a 3d ${\cal N}=2$ theory. In this case, there is a simple M-theory interpretation - one has precisely the theory of M5 branes wrapping $M_3 \times \mathbb{R}^{2,1}$ in
the eleven-dimensional supersymmetric geometry $T^* M_3 \times \mathbb{R}^{4,1}$.

There exist many other examples of supersymmetric curved backgrounds \cite{Nekrasov:2002qd, Pestun:2007rz, Kapustin:2009kz, Jafferis:2010un, Hama:2010av, Dumitrescu:2012ha}.  In particular, the partition function of 3d SCFTs on squashed spheres has proven to be a very useful characteristic quantity. It can be computed exactly for theories with a Lagrangian description using supersymmetric localization. The logarithm of the round sphere partition function is an monotonic measure of the number of degrees of freedom\cite{Jafferis:2010un, Jafferis:2011zi, Myers:2010xs, Myers:2010tj, Casini:2011kv, Casini:2012ei}. The dependence of the partition function on a supersymmetric squashing parameter encodes further information such as the two point function of the R current \cite{Closset:2012ru}.

We will be interested in the partition function on the squashed three-sphere, $S^3_\ell$, described in \cite{Imamura:2011wg}, which preserves an $SU(2) \times U(1)$ isometry, where $\ell/2$ is the ratio of the radius of the $S^2$ base and the $S^1$ Hopf fiber. The round sphere is given by $\ell = 1$. Another geometry also sometimes know as the squashed sphere is the ellipsoid, $S^3_b$, which preserves $U(1) \times U(1)$ isometry \cite{Hama:2011ea}. The ${\cal N}=2$ partition functions on these geometries are equal with the identification $\ell = \frac{2}{b + b^{-1}}$. Note that this only covers the range $0 < \ell \leq 1$ for which the Hopf fiber is larger than in the round sphere. 

The $S^3_{\ell}$ partition function of the field theory $T_\frak{g}(M_3)$ is a quantity which does not depend on the metric on $M_{3}$ and hence must be a topological invariant of the three-manifold. It was conjectured by \cite{Ooguri:1999bv, Terashima:2011qi, Terashima:2011xe, Dimofte:2011ju, Cecotti:2011iy, Dimofte:2011py, Galakhov:2012hy,  Beem:2012mb, Yagi:2013fda, Gang:2013sqa} to be related to the partition function of pure Chern-Simons theory with a noncompact gauge group on $M_3$. In many ways, this is surprising, since naively a supersymmetric reduction of the 6d (2,0) theory would result in a supersymmetric theory with a compact gauge group, not a non-supersymmetric theory with a noncompact gauge group. This relationship is similar in spirit to the AGT conjecture \cite{Alday:2009aq, Alday:2009fs} which equates instanton partition functions of supersymmetric four-dimensional field theories with non-supersymmetric Toda conformal blocks.

In principle there is nothing mysterious about deriving such a hypothetical 3d-3d correspondence. Consider the $(2,0)$ theory on $S^3_\ell \times M_3$, partly topologically twisted on $M_3$ and with the appropriate background fields to preserve supersymmetry on $S^3_\ell$. The supersymmetric partition function is independent of the relative size of the sphere and three-manifold due to the twisting - all dependence on the three-manifold metric is $Q$-exact. It also does not depend on the overall scale due to conformal invariance of the $(2,0)$ theory.\footnote{The Euler character and other relevant
invariants vanish in this case, so the 6d conformal anomalies do not contribute.}

If there had been a Lagrangian for the 6d theory, one would simply reduce on a small $M_3$ to obtain a 3d ${\cal N}=2$ Lagrangian theory which flows to the SCFT $T_\frak{g}(M_3)$ in the IR, and whose $S^3_\ell$ partition function could be computed directly.  On the other hand, one could also consider the reduction in the opposite limit of a small sphere. In that limit, one would again Kaluza-Klein reduce to obtain another 3d Lagrangian theory, parameterized by $\ell$, whose partition function on $M_3$ should give the same result.  The obstruction to carrying such a procedure is that no Lagrangian for the 6d $(2,0)$ theory currently exists.  This fact is one of the reasons why it is interesting to study the theories $T_\frak{g}(M_3)$, since these compactifications provide one of the few windows into a broader class of 3d  ${\cal N}=2$ theories.

In this paper we bypass these difficulties by noting that the squashed sphere geometry is simple, and always possesses a circle isometry. Therefore the reduction on $S^3_\ell$ can proceed in two stages. First, we reduce along the circle, leading to 5d maximally supersymmetric Yang-Mills \cite{Seiberg:1997ax} with gauge group $\mathfrak{g}$ in the background geometry $S^2 \times M_3$, and coupled to various background fields required by supersymmetry. Then, we reduce on the $S^2$ the resulting Lagrangian field theory.  Related construction have appeared in \cite{Fukuda:2012jr, Kawano:2012up, Yagi:2013fda, Kim:2012qf}.

In general, the dimensional reduction of the $(2,0)$ theory on a circle only looks like 5d Yang-Mills at low energies. There will be an additional series of higher order irrelevant operators that become important as the coupling of the 5d non-renormalizable theory grows at high energies \cite{Douglas:2010iu, Lambert:2010iw, Kim:2011mv, Bonetti:2012st, Bern:2012di}. The 5d Yang-Mills coupling is given by the size of the Hopf circle, and we are interested in the dimensional reduction limit where the Hopf fiber, and indeed the entire $S^{3}_{\ell}$ geometry has vanishing size.  Therefore our interest is with the weakly coupled limit of 5d Yang-Mills where all higher derivative corrections arising from the reduction of the 6d $(2,0)$ theory are parametrically suppressed.\footnote{The 5d theory does look strongly coupled at the scale of the $S^2$, however only the zero mode sectors will contribute to the 3d effective action, so higher order operators are not expected to affect the answer. Moreover, in a supersymmetric partition function such as the one we wish to compute, it is a reasonable conjecture that all higher order operators are $Q$-exact and do not change the exact partition function.}

The intermediate step of 5d Yang-Mills is particularly important in the non-abelian theory.   As there is no 6d Lagrangian, one must instead proceed on general grounds to find the coupling of Yang-Mills to off-shell supergravity.  Such a procedure is possible because of the maximal supersymmetry in 5d which implies that the coupling to supergravity is unique; there is no choice in the multiplet in which the stress tensor sits.  Moreover, in the squashed sphere background, a background term adds a cubic potential for the scalars in the Yang-Mills multiplet, whose six dimensional origin is obscure.

Having arrived in 5d Yang-Mills, an ordinary Lagrangian field theory, it would now seem that further dimensional reduction on $S^{2}$ must produce a supersymmetric $\mathfrak{g}$ gauge theory, something completely different than bosonic $\mathfrak{g}_{\mathbb{C}}$ Chern-Simons. Amazingly, it will turn out that the two seeming problems, lack of supersymmetry and complexification of the gauge group, cancel each other. The fermionic superpartners of the gauge field will turn into Faddeev-Popov ghosts of partial gauge fixing of $\mathfrak{g}_{\mathbb{C}}$ to $\mathfrak{g}$. In this way, one obtains $\mathfrak{g}_{\mathbb{C}}$ Chern-Simons theory at levels $k,$ and $u$ stated in \eqref{leveldefintro}. We conclude that the squashed three-sphere partition function of the ${\cal N}=2$ 3d SCFT given by compactification of the $\mathfrak{g}$ $(2,0)$ theory on a 3-manifold, $M_3$, is exactly equal to the partition function of $\mathfrak{g}_{\mathbb{C}}$ Chern-Simons on $M_3$
\begin{equation} 
Z_{S^3_\ell}[T_\frak{g}(M_3)] = Z_{M_3}[CS_{\frak{g}_C}(1, \sqrt{1-\ell^2})].
\end{equation}

Let us now describe in more detail the structure of the arguments to follow.
\begin{comment}
We begin in section \ref{sugra} with off-shell $(2,0)$ supergravity in six dimensions.  This theory was constructed in \cite{Bergshoeff:1999db}.  Since we are interested in treating those fields as classical backgrounds, all fermions in the supergravity multiplet will be set to zero. The vanishing of the gravitini and dilatini supersymmetry variations for particular spinor parameters will imply that the coupled dynamical theory has an associated rigid supersymmetry.
\end{comment}

The six dimensional background of interest is $S^3_\ell \times M_3$, preserving $SU(2|1) \times U(1)$ supersymmetry.  It is a non-trivial fact that such a configuration preserving four supercharges exists in six dimensions. The case of the round sphere may be understood in a simple way. The geometry $H_3 \times S^3$ with equal radii is conformally flat. Therefore, one may conformally map the $(2,0)$ theory to this space, preserving full superconformal invariance. It is then possible to twist $H_3$ to a general three-manifold metric, preserving four supercharges. The squashed sphere case is more involved, and also involves the anti-self-dual 3-form of in the 6d supergravity multiplet.

In order to have an action, we reduce along the circle fiber of the squashed sphere to maximal 5d super-Yang-Mills coupled to 5d supergravity background fields.  The conditions for preserving supersymmetry in a configuration of background maximal 5d supergravity were found in \cite{us}.   Since we are interested in a 5d calculation, in section \ref{backgroundfields} we describe the squashed $S^{3}_{\ell}$ geometry from the point of view of 5d supergravity.  We find that all background fields must be used in the $S^2 \times M_3$ reduction at general squashing. The squashing parameter controls the ratio of the dilaton and the $S^2$ radius, as well as other background fields, as required by supersymmetry.  There is a single unit of graviphoton flux wrapping the $S^{2}$ which results from the fact that the Hopf fibration is non-trivial.  

Next, in section \ref{abelianzeromodes}, we dimensionally reduce 5d Yang-Mills in the given supergravity background.  As our construction is compatible with topological twisting on a general $M_{3}$ it is sufficient to consider the five-dimensional geometry $S^2\times \mathbb{R}^{3}$.  The resulting 3d theory can then be placed on any 
three-manifold manifold $M_{3}$ if desired. We find that various background fields are activated in $\mathbb{R}^{3}$, and one must include their contributions in finding the effective action.

We obtain the following light fields in 3d.  
\begin{itemize}
\item The constant mode of the 3d components of the gauge field $A$.  

The 5d graviphoton-Chern-Simons coupling produces a 3d Chern-Simons term at level one for this field.
\begin{equation}
\frac{1}{8\pi^{2}}\int_{\mathbb{R}^{3}\times S^{2}} \hspace{-.15in}C \wedge \mathrm{Tr}\left(F\wedge F\right) \rightarrow \frac{1}{4\pi} \int_{\mathbb{R}^{3}}\mathrm{Tr}\left(A\wedge dA+\frac{2}{3}A\wedge A \wedge A\right).
\end{equation}
The remaining Yang-Mills terms vanish in the $r\rightarrow0$ compactification limit.  This is responsible for the $\mathfrak{g}$ sector of our result with the indicated value of the level $k=1$.

\item The the zero modes of the five scalars $\varphi_{\hat{D}}$, with $\hat{D}=1, ..., 5$ an $so(5)_{R}$ symmetry index.  Under the preserved $so(3)$ symmetry, identified as the Lorentz symmetry of the resulting 3d theory, the five scalars are naturally split into a triplet $X_{a}$, and a pair of singlets $Y_{i}$.  Thus it is natural to interpret $X_a$ as a 1-form. But note that it is not (yet) associated to any gauge symmetry.  

In addition to the usual interaction terms that occur in flat space, the action for the fields $X_{a}$ contains interesting terms which arise from the coupling to supergravity.
 \begin{itemize}
\item The R gauge field $V$ is activated inside an $so(3) \subset so(5)_{R}$.  The 5d kinetic term for the fields $\varphi_{\hat{D}}$ then produces a term which is first order in derivatives for the $X$'s
\begin{equation}
\mathrm{Tr}\left(V_{a}\varphi \nabla^{A}_{a}\varphi\right)\rightarrow\varepsilon^{abc}\mathrm{Tr}\left(X_{a}\nabla^{A}_{b}X_{c}\right).  \label{term1}
\end{equation}
\item There is a 2-from $T^{\hat{B}}$ in the $\mathbf{5}$ of $so(5)_{R}$.  This field is activated inside $so(3) \subset so(5)_{R}$. It results in the following term
\begin{equation}
\mathrm{Tr}\left(\varphi^{\hat{B}}F\wedge * T_{\hat{B}}\right) \rightarrow \sqrt{1-\ell^{2}}\varepsilon^{abc}\mathrm{Tr}\left(X_{a}F_{bc} \right).\label{term2}
\end{equation} 
\item There is an R-scalar $S_{\hat{A}\hat{B}}$ in the adjoint of $so(5)_{R}$.  This field is activated inside $so(2)\subset so(5)_{R}$.  It generates a cubic potential for the scalars
\begin{equation}
\varepsilon^{\hat{A}\hat{B}\hat{C}\hat{D}\hat{E}}S_{\hat{A}\hat{B}}\mathrm{Tr}\left(\varphi_{\hat{C}}[\varphi_{\hat{D}},\varphi_{\hat{E}}]\right)\rightarrow  -\frac{\sqrt{1-\ell^{2}}}{3}\varepsilon^{abc}\mathrm{Tr}\left(X_{a}X_{b}X_{c}\right).\label{term3}
\end{equation}
\end{itemize}
The interactions, \eqref{term1}-\eqref{term3}, produced automatically by the coupling to supergravity, are responsible for the additional terms in the complex Chern-Simons action \eqref{SCS} together with the indicated value of the parameter $u=\sqrt{1-\ell^{2}}$.
\item Four fermions $\lambda$ associated to the four preserved supersymmetries of the compactification.  
\end{itemize}
In section \ref{rzero} we complete the analysis of the resulting theory of these zero modes.  

The massless fermionic action has an interesting peculiarity - it appears to vanish identically. This is because the fermionic kinetic terms are not diagonalized in the basis of modes which diagonalizes the mass matrix. Thus one needs to initially keep all of the fermionic modes that can mix with massless ones, and then integrate them out. This will result in a fermionic effective action that is second order in derivatives.

The full action for the $Y$ and $\lambda$ fields takes the qualitative form
\begin{equation}
r\int d^{3}x \ \mathrm{Tr}\left(\phantom{\int}\hspace{-.17in}-\lambda\left(\nabla^{A}\right)^{2}\lambda-Y\left(\nabla^{A}\right)^{2}Y+[X,\lambda]^{2}+[X,Y]^{2}+[Y,Y]^{2}+[Y,\lambda]^{2}\right). \label{ssketch}
\end{equation}
This action takes a form very similar to that which appears in Faddeev-Popov gauge fixing, with the fermions playing the role of the ghosts for a gauge fixing term $\nabla^{A}_{b} X^{b} = 0$ of the gauge transformation $X_b \rightarrow X_b + \nabla^A_b g, \ A_c\rightarrow A_c - [X_c, g]$ for local gauge parameter $g$. However, there appear to be twice as many fermions as required, and there are additional scalars, $Y$, as well as non-linear interaction terms.

To understand the action \eqref{ssketch}, we observe that the preserved supersymmetries are given schematically as
 \begin{eqnarray}
\delta A & = & \nabla^{A}\left(\beta\lambda\right), \nonumber \\
\delta X & = & [X,\beta\lambda] ,\\
\delta Y_{k} & = & [Y_{k},\beta \lambda]+i[\varepsilon_{kj}Y_{j},\beta \lambda], \nonumber \\
\delta \lambda & = & [Y_{1},Y_{2}]\beta, \nonumber
\end{eqnarray}
where in the above $\beta$ is a Grassmann parameter.  Interestingly, the action of these supersymmetries has become almost trivial (up to a gauge transformation) in the small $r$ limit, acting only on the fermions and $Y$.  It follows that supersymmetric observables which are functions only of the fields $X$ and $A$ coincide with the gauge invariant observables.  Provided that we restrict ourselves to this class of expectation values, we are free to deform the action by a $Q$-exact term of the form
\begin{equation}
\delta \tr(\lambda [Y,Y]) = \tr( [Y,Y][Y,Y] + [\lambda, Y][Y, \lambda]).
\end{equation}
This allows us to remove all non-linear terms in the action for $Y$ and $\lambda$. 

The resulting quadratic action for $Y$ has a 1-loop determinant that is identical to that of the fermions. Therefore the exact path integral over the $\lambda$ and $Y$ fields produces exactly the Faddeev-Popov determinant for the gauge fixing term! The $r
\rightarrow 0$ limit corresponds to a singular choice of gauge, and can now be taken by simply undoing the gauge fixing.

Finally, we summarize in section \ref{discussion}.  Details of the 5d action and the relation between 6d and 5d background fields may be found in appendix \ref{5dsymdetails}. The relevant spinor algebra in collected in appendix \ref{cliffordcovnetions}.
\section{The M5-Brane on a Squashed Three-Sphere}
\label{backgroundfields}
In this section, we construct a Euclidean continuation of the six-dimensional $(2,0)$ theory on a three-dimensional spherical background.  The total space of the geometry is $\mathbb{R}^{3}\times S^{3}_{\ell}$ where, topologically $S^{3}_{\ell}$ is a three-dimensional sphere.  We view $S^{3}_{\ell}$ as a Hopf fibration.  
\begin{equation}
\xymatrix{S^{1}\ar[r] & S^{3}_{\ell} \ar[d]\\
& S^{2}}
\end{equation}
The round metric preserving an $su(2)\times su(2)$ isometry is achieved when the ratio of the radius of the base $S^{2}$ to the radius of the fiber $S^{1}$ is $1/2$.   More generally, we consider a squashed three-sphere which preserves the round form of the metrics on the base and the fiber, but takes their ratio of radii to be $\ell/2$, where $\ell\in \mathbb{R}_{+}$ is a parameter of the construction.  For a generic value of $\ell$ the squashed three-sphere has an $su(2)\times u(1)$ isometry algebra, realized as rotations of the base and fiber respectively.  

Preserving supersymmetry on a squashed three sphere will require us to activate various background supergravity fields,  and in this section we describe the precise form of these fields.  Throughout we make use of the $U(1)$ isometry group which rotates the Hopf fiber to reduce the calculation to one in five-dimensional supergravity.  

The fields of the off-shell 5d sueprgravity multiplet are enumerated in Table \ref{5dgravfields}, including their R symmetry representations and scaling dimension $w$.  The possible backgrounds consist of arbitrary configurations of the bosonic fields and vanishing profiles of the fermionic fields.   The background is supersymmetric provided that the variation of the fermions vanishes
\begin{equation}
\delta \psi= \delta \chi =0.
\end{equation}
The precise form of these variations are given in appendix \ref{5dsymdetails}.

In our problem the choice of background fields is dictated by symmetry.  In particular, we wish to study a configuration which is compatible with replacing $\mathbb{R}^{3}$ with a general three-manifold $M_{3}$ while preserving supersymmetry.  It follows that our choice of background supergravity fields must be compatible with topological twisting.  We find that for a suitable choice of background fields, four total supercharges can be preserved, while respecting all the symmetries required by $\mathcal{N}=2$ twisting on a general $M_{3}$.\footnote{A distinct compactification preserving two total supercharges related to $\mathcal{N}=1$ supersymmetry in three dimensions is also possible and will be analyzed elsewhere.}  A summary of the backgrounds is given in section \ref{sugraconf}, more detailed calculations may be found in section \ref{backcal}.
\begin{table}[h]
\centering
\begin{tabular}{|c|c|c|c|c|}
\hline
Field & Type & Name and Properties & $so(5)_{R} $ & w \\
\hline
\multirow{2}{*}{$e^{a}_{\mu}$} & \multirow{2}{*}{boson} & \multirow{2}{*}{coframe} &\multirow{2}{*}{$\mathbf{1}$} & \multirow{2}{*}{-1}\\
 & &  & & \\
 \hline
 \multirow{2}{*}{$C_{\mu}$} & \multirow{2}{*}{boson} & \multirow{2}{*}{graviphoton \hspace{.2in}$G\equiv dC$} &\multirow{2}{*}{$\mathbf{1}$} & \multirow{2}{*}{0}\\
 & &  & & \\
\hline
 \multirow{2}{*}{$\alpha$} & \multirow{2}{*}{boson} & \multirow{2}{*}{dilaton} &\multirow{2}{*}{$\mathbf{1}$} & \multirow{2}{*}{1}\\
 & &  & & \\
\hline
 \multirow{2}{*}{$V_{\mu}^{\hat{A}\hat{B}}$} & \multirow{2}{*}{boson} & \multirow{2}{*}{R gauge field \hspace{.3in} $V_{\mu}^{\hat{A}\hat{B}}=-V_{\mu}^{\hat{B}\hat{A}}$} &\multirow{2}{*}{$\mathbf{10}$} & \multirow{2}{*}{0}\\
 & &  & & \\
\hline
\multirow{2}{*}{$S^{\hat{A}\hat{B}}$} & \multirow{2}{*}{boson} & \multirow{2}{*}{$S^{\hat{A}\hat{B}}=-S^{\hat{B}\hat{A}}$} &\multirow{2}{*}{$\mathbf{10}$} & \multirow{2}{*}{1}\\
 & &  & & \\
\hline
\multirow{2}{*}{$ T^{\hat{A}}_{\mu\nu}$} & \multirow{2}{*}{boson} &\multirow{2}{*}{$T^{\hat{A}}_{\mu \nu}=-T^{\hat{A}}_{\nu \mu},$} & \multirow{2}{*}{$\mathbf{5}$} & \multirow{2}{*}{-1} \\
 & &  & &\\
\hline
\multirow{2}{*}{$ D^{\hat{A},\hat{B}}$} & \multirow{2}{*}{boson} &  \multirow{2}{*}{$D^{\hat{A},\hat{B}}=D^{\hat{B},\hat{A}}, \hspace{.5in}\delta_{\hat{A}\hat{B}}D^{\hat{A}\hat{B}}=0$}& \multirow{2}{*}{$\mathbf{14}$} & \multirow{2}{*}{2} \\
 & &  & &\\
\hline
\multirow{2}{*}{$\psi^{m}_{\mu}$} & \multirow{2}{*}{fermion} & \multirow{2}{*}{symplectic Majorana ``gravitini''} &\multirow{2}{*}{$\mathbf{4}$} & \multirow{2}{*}{$-1/2$}\\
 & &  & & \\
\hline
\multirow{2}{*}{$ \chi^{ mn }_{r}$} &\multirow{2}{*}{fermion} & $\chi^{mn}_{r}=-\chi^{nm}_{r}, \hspace{.5in} \Omega_{mn}\chi^{mn}_{r}=\delta^{r}_{m}\chi^{mn}_{r}=0,$& \multirow{2}{*}{$\mathbf{16}$} & \multirow{2}{*}{3/2} \\
& & symplectic Majorana  ``dilatini'' & & \\
\hline
\end{tabular}
\caption{Fields of five-dimensional off-shell $\mathcal{N}=2$ supergravity.}
\label{5dgravfields}
\end{table}
\subsection{Supergravity Configuration}
\label{sugraconf}
We begin with a symmetry analysis.  In five dimensions the Lorentz and R symmetry algebra is $so(5)_{L}\times so(5)_{R}$.  To describe our background we first split this group as 
\begin{equation}
so(5)_{L}\times so(5)_{R}\longrightarrow so(2)_{L}\times so(3)_{L} \times so(3)_{R}\times so(2)_{R}.
\end{equation}
The group $so(2)_{L}$ is realized geometrically as a group of rotations of the sphere $S^{2}$, while the group $so(2)_{R}$ is a global R symmetry of the theory.  Finally, the group $so(3)_{L}\times so(3)_{R}$ is broken to its diagonal subgroup.  Note that this is the minimal symmetry group compatible with further topologically twisting the theory on a general three manifold other than $\mathbb{R}^{3}$.

To write the fields explicitly we use the following index conventions
\begin{itemize}
\item $so(5)_{L}$ vector indices are indicated by capital Roman letters $A, B, \cdots.$ 
\item $so(3)_{L}$ vector indices are indicated by lowercase Roman letters from the first few letters of the alphabet $a, b \cdots.$ 
\item $so(2)_{L}$ vector indices are indicated by lowercase Roman letters from the last few letters of the alphabet $w, x \cdots.$ 
\item $so(3)_{L}$ spinor indices are indicated by lowercase Greek letters from the first half of the alphabet $\alpha, \beta \cdots.$
\item $so(2)_{L}$ spinor indices are indicated by lowercase Greek letters from the second half of the alphabet $\sigma, \tau \cdots.$ 
\item An R-symmetry index is distinguished from a Lorentz index by adding a hat.  For example, $\hat{\alpha}$ indicates an $so(3)_{R}$ spinor index.
\item $so(5)_{R}$ spinor indices are indicated by lowercase Roman letters from the middle of the alphabet $m, n, \cdots.$
\end{itemize}

After these preliminaries we now enumerate the background fields.  The metric of the squashed three-sphere is encoded in the five-dimensional metric, graviphoton and dilaton.  These are given by
\begin{equation}
ds^{2}=dx_{0}^{2}+dx_{1}^{2}+dx_{2}^{2}+\left(\frac{r\ell}{2}\right)^{2}\left(d\theta^{2}+\sin^{2}(\theta)d\phi^{2}\right), \hspace{.5in} C=\cos^{2}(\theta/2)d\phi, \hspace{.5in} \alpha=1/r.
\end{equation}
The case $\ell=1$ describes the round three-sphere.  For notational convenience, we find it useful to define the following quantity
\begin{equation}
e\equiv\sqrt{\frac{1-\ell}{1+\ell}},
\end{equation}
which appears in various calculations below.

The profiles of the remaining fields in the supergravity multiplet are greatly constrained by symmetry.  They take the form
\begin{equation}
T_{\hat{A}BC}=t\varepsilon_{\hat{a}bc}, \hspace{.5in}V_{A\hat{B}\hat{C}}= v \varepsilon_{a\hat{b}\hat{c}},\hspace{.5in} S_{\hat{A}\hat{B}}=s\varepsilon_{\hat{x}\hat{y}},\hspace{.5in} D_{\hat{A}\hat{B}}=d\left(\delta_{\hat{a}\hat{b}}-\frac{3}{2}\delta_{\hat{x}\hat{y}}\right). \label{ansatz}
\end{equation}
In the above $t,v,s,d$ are constants.  They must be determined by requiring that the variations of supergravity fermions $\delta \psi$ and $\delta \chi$ vanish.   Note that the R gauge field $V$ vanishes in the two-sphere directions.  In principle a more general ansatz consistent with the symmetries would allow a non-vanishing $V$ on the $S^{2}$ whose two-from field strength was proportional to the volume form in the round metric.  Flux quantization, would then yield quantization of the field strength.  We find that it is consistent with supersymmetry to set this quantized flux parameter to zero.  We may then reach a gauge where the background fields take the form described by \eqref{ansatz}

The fields stated above are given in terms of their $so(5)_{R}$ representation content.  As the supersymmetry parameters  $\epsilon^{m}$ are in the spinor representation of this group, it is useful to convert to $sp(4)_{R}$ representations as follows
\begin{equation}
\begin{tabular}{l l l l l l}
$T^{mn}_{BC}=T_{\hat{A}BC}\left(\Gamma^{\hat{A}}\right)^{mn},$ & & & & &$V_{An}^{\phantom{A}m} =V_{A\hat{B}\hat{C}}\left(\Gamma^{\hat{B}\hat{C}}\right)^{m}_{n},$ \\ 
\\
$S^{m}_{n}=S_{\hat{A}\hat{B}}\left(\Gamma^{\hat{B}\hat{C}}\right)^{m}_{n},$  & & & & &$D^{mn,rs}=D_{\hat{A}\hat{B}}\left(\Gamma^{\hat{A}}\right)^{mn}\left(\Gamma^{\hat{B}}\right)^{rs}.$
\end{tabular}
\label{so5sp4}
\end{equation}
Under the reduction of $so(5)_{R}$ to $so(3)_{R}\times so(2)_{R}$, the $sp(4)$ invariant tensors decompose into products of invariant tensors given in appendix \ref{32facts} as 
\begin{equation}
\delta^{m}_{n}\rightarrow \delta^{\hat{\alpha}}_{\hat{\beta}}\delta^{\hat{\sigma}}_{\hat{\tau}}, \hspace{.5in} \Omega_{mn}\rightarrow \varepsilon_{\hat{\alpha}\hat{\beta}}B_{\hat{\sigma}\hat{\tau}}.
\end{equation}

Finally we also require that the spinor parameter $\epsilon$ generating supersymmetry transformations transforms as a singlet under the unbroken $so(3)$ symmetry group and is constant in $\mathbb{R}^{3}$
\begin{equation}
\epsilon^{\alpha \sigma \hat{\alpha}\hat{\sigma}}= \varepsilon^{\alpha \hat{\alpha}}\eta^{\sigma \hat{\sigma}}, \hspace{.5in}\partial_{a}\epsilon^{\alpha \sigma \hat{\alpha}\hat{\sigma}}=0. \label{spinorfactorized}
\end{equation}
These properties ensure that our construction is compatible with topological twisting, and allow us to immediately generalize the construction from $\mathbb{R}^{3}$ to a general three manifold.

A direct calculation outlined in section \ref{backcal} using this ansatz determines the values of the parameters required for supersymmetry.  We find
\begin{equation}
t=s=-\frac{\sqrt{1-\ell^{2}}}{2r\ell^{2}}, \hspace{.5in}v =  -\frac{i}{2r\ell^{2}}, \hspace{.5in} d=\frac{3}{2r^{2}\ell^{2}}\left(1+\frac{1}{\ell^{2}}\right). \label{constants}
\end{equation}
For these values, four total supercharges are preserved.
\subsection{Analysis of Supersymmetry Constraints}
\label{backcal}

To demonstrate that our ansatz describes a supersymmetric background, we must show that there exist non-zero spinors $\epsilon^{m}$ with the property that
\begin{equation}
\delta \psi^{m}=\delta \chi^{mn}_{r}=0,
\end{equation}
where the above variations are defined by \eqref{5dsusyvar}.  We solve these equations by separating the spinors $\epsilon^{m}$ into a tensor product of spinors on $\mathbb{R}^{3}$ and spinors on $S^{2}$ as indicated in \eqref{spinorfactorized} and described in detail in appendix \ref{5=3+2} .

\subsubsection{$\delta \psi_{x}^{m}$,  $x\in S^{2}$}
The non-vanishing contributions are
\begin{equation}
\delta \psi_{x}^{m}=\left(\partial_{x}+\frac{1}{4}\omega_{x}^{yz}\Gamma_{yz}\right)\epsilon^{m}+\frac{i}{2\alpha}G_{xy}\Gamma^{y}\epsilon^{m}-\frac{i}{2}S^{mn}\Gamma_{x}\epsilon_{n}-\frac{i}{2}T^{mnbc}\Gamma_{xbc}\epsilon_{n}=0
\end{equation}
We simplify the above using our ansatz for the background fields.  We find linear differential equations of the form
\begin{eqnarray}
\partial_{\theta}\eta^{\sigma \hat{\sigma}} &= & \left[\frac{i}{2\ell}\left(\kappa^{4}\right)^{\sigma}_{\tau}\delta^{\hat{\sigma}}_{\hat{\tau}}+\frac{r\ell(3t-s)}{2}\left(\kappa^{3}\right)^{\sigma}_{\tau}\kappa^{\hat{\sigma}}_{\hat{\tau}}\right]\eta^{\tau \hat{\tau}}, \label{firstsquashed} \\
\partial_{\phi}\eta^{\sigma \hat{\sigma}} & = & \left[    \frac{r\ell (3t-s)\sin(\theta)}{2}\left(\kappa^{4}\right)^{\sigma}_{\tau}\kappa^{\hat{\sigma}}_{\hat{\tau}} -\frac{i\cos(\theta)}{2}\kappa^{\sigma}_{\tau}\delta^{\hat{\sigma}}_{\hat{\tau}}-\frac{i\sin(\theta)}{2\ell}\left(\kappa^{3}\right)^{\sigma}_{\tau}\delta^{\hat{\sigma}}_{\hat{\tau}}\right] \eta^{\tau \hat{\tau}}, \nonumber
\end{eqnarray}
where in the above, $\kappa^{3}, \kappa^{4}$ and $\kappa$ are the two-dimensional Clifford algebra and chirality matrix defined in equation \eqref{2dcliffdefs}, and we have used the fact that the non-vanishing components of the spin connection on the two-sphere take the form $\omega_{\phi}^{43}=-\omega_{\phi}^{34}=\cos(\theta).$

To solve these equations we note that by iteration we obtain
\begin{eqnarray}
\partial^{2}_{\theta}\eta^{\sigma \hat{\sigma}} &= & \left[\frac{r^{2}\ell^{2}(3t-s)^{2}}{4}-\frac{1}{4\ell^{2}}\right]\eta^{\sigma \hat{\sigma}} \label{iteratedsquashed}\\ 
\partial^{2}_{\phi}\eta^{\sigma \hat{\sigma}} & = &\left[\left(\frac{r^{2}\ell^{2}(3t-s)^{2}}{4}-\frac{1}{4\ell^{2}}\right)\sin^{2}(\theta)-\frac{\cos^{2}(\theta)}{4}\right]\eta^{\sigma \hat{\sigma}}. \nonumber
\end{eqnarray}
Hence, if the values of $s$ and $t$ satisfy 
\begin{equation}
s-3t=\frac{\sqrt{1-\ell^{2}}}{r\ell^{2}},
\end{equation}
we see that \eqref{iteratedsquashed} simplifies to
\begin{equation}
\partial^{2}_{\theta}\eta^{\sigma \hat{\sigma}}=-\frac{1}{4}\eta^{\sigma \hat{\sigma}}, \hspace{.5in}\partial^{2}_{\phi}\eta^{\sigma \hat{\sigma}}=-\frac{1}{4}\eta^{\sigma \hat{\sigma}}.
\end{equation}
For this choice of parameters the first order equations \eqref{firstsquashed} may be solved to yield four linearly independent solutions.  These take the form
\begin{equation}
\begin{tabular}{lllll}
$\eta_{++}\equiv\frac{\exp\left(\frac{i\phi}{2}\right)}{\sqrt{4\pi r\ell^{2}}}\left[\begin{array}{cc}(1+e)i\sin(\theta/2)  &  0\\  (1-e)\cos(\theta/2) & 0\end{array}\right],$ & &&& $\eta_{+-}\equiv\frac{\exp\left(\frac{i\phi}{2}\right)}{\sqrt{4\pi r\ell^{2}}}\left[\begin{array}{cc}0  &(1-e)i\sin(\theta/2)  \\ 0 &  (1+e)\cos(\theta/2) \end{array}\right],$ \\
\\
$\eta_{-+}\equiv\frac{\exp\left(\frac{-i\phi}{2}\right)}{\sqrt{4\pi r\ell^{2}}}\left[\begin{array}{cc}(1+e)\cos(\theta/2)   & 0\\  (1-e)i\sin(\theta/2) &0\end{array}\right],$ & &&& $\eta_{--}\equiv\frac{\exp\left(\frac{-i\phi}{2}\right)}{\sqrt{4\pi r\ell^{2}}}\left[\begin{array}{cc}0 & (1-e)\cos(\theta/2)  \\ 0 &  (1+e)i\sin(\theta/2) \end{array}\right],$
\end{tabular}
\label{squashedspinors}
\end{equation}
where in the above our convention is that the $\sigma$ index labels rows and the $\hat{\sigma}$ index labels columns.  The overall normalization of the solutions is arbitrary and chosen for later convenience.  A general solution thus takes the form
\begin{equation}
\eta^{\sigma \hat{\sigma}}=\beta^{i\hat{i}}\eta_{i\hat{i}}^{\sigma \hat{\sigma}},
\end{equation}
with $\beta^{i\hat{i}}$ Grassmann coefficients.

\subsubsection{$\delta \psi_{a}^{m}$,  $a\in \mathbb{R}^{3}$}
The non-vanishing contributions are 
\begin{equation}
\delta \psi_{a}^{m}=-\frac{1}{2}V_{an}^{\phantom{a}m}\epsilon^{n}-\frac{i}{2}S^{mn}\Gamma_{a}\epsilon_{n}+\frac{i}{8\alpha}G^{xy}\Gamma_{axy}\epsilon^{m}-\frac{i}{2}T^{mnbc}\Gamma_{abc}\epsilon_{n}=0
\end{equation}
Simplifying using our ansatz for the background fields, the above reduces to
\begin{equation}
\left(\frac{1}{2r\ell^{2}}-iv\right)\eta^{\sigma \hat{\sigma}}+(t-s)\kappa^{\sigma}_{\tau}\kappa^{\hat{\sigma}}_{\hat{\tau}}\eta^{\tau \hat{\tau}}=0. \label{r3squashed}
\end{equation}
Given the solutions to equation \eqref{squashedspinors}, non-trivial solutions to the above can only be obtained if
\begin{equation}
s=t=-\frac{\sqrt{1-\ell^{2}}}{2r\ell^{2}}, \hspace{.5in} v=-\frac{i}{2r\ell^{2}}.
\end{equation}
For this value of background field, the four supercharges found on the two sphere are consistent with $\delta \psi_{A}=0$ for all values of the index $A$.

\subsubsection{$\delta \chi^{mn}_{r}$}
We turn to the variation of $\chi^{mn}_{r}.$  For simplicity, we present the calculation in the special case of the round three-sphere when $\ell=1$.  The analysis at general values of the squashing parameter is similar but technically more involved.

In the special case of the round three-sphere, the background values of $s$ and $t$ vanish.  According to formula \eqref{5dsusyvar}, this implies that the variation of $\chi^{mn}_{r}$ receives non-vanishing contributions from the curvature of the R gauge field and the scalar $D$ field.  Including all traces explicitly, the variation takes the form
\begin{equation}
\delta \chi ^{mn}_{r}=\left(\frac{1}{5}R_{abs}^{\phantom{a}[m}\delta_{r}^{n]}+\frac{1}{5}R_{abrs}\Omega^{mn}-R_{abr}^{\phantom{a}[m}\delta^{n]}_{s}\right)\Gamma^{ab}\epsilon^{s}-\frac{4}{15}D^{mn}_{rs}\epsilon^{s}. \label{dilatinov}
\end{equation}
We now evaluate the above using the following results derived from our analysis of $\delta \psi$.  We write all terms in the reduced $so(3)_{R}\times so(2)_{R}$ index form with the labeling conventions
\begin{equation}
m\rightarrow (\hat{\alpha},\hat{\sigma}),\hspace{.4in}n\rightarrow (\hat{\beta},\hat{\tau}),\hspace{.4in}r\rightarrow (\hat{\gamma},\hat{\upsilon}),\hspace{.4in}s\rightarrow (\hat{\zeta},\hat{\omega}).
\end{equation}
We similarly split the $so(5)_{L}$ Lorentz spinor indices in to $so(3)_{L}\times so(2)_{L}$ pairs.  For example,
\begin{equation}
\Gamma^{ab}\varepsilon \rightarrow \left(\Gamma^{ab}\right)^{\alpha \sigma}_{\beta \tau}\epsilon^{\beta \tau}.
\end{equation}

The non-vanishing elements of the curvature tensor are calculated from \eqref{5dcurvdefs} and \eqref{ansatz} and our explicit value of $v$ in \eqref{constants}.  They take the form
\begin{equation}
R_{abn}^{\phantom{ab}m}=-\frac{1}{r^{2}} (\gamma_{ab})^{\hat{\alpha}}_{\hat{\beta}}\delta^{\hat{\sigma}}_{\hat{\tau}}. 
\end{equation}
From the above we deduce that the contraction $R_{abn}^{\phantom{ab}m}\Gamma^{ab}$ can be written in reduced index form as
\begin{equation}
\slashed{R}^{\alpha \sigma \hat{\alpha}\hat{\sigma}}_{\beta \tau \hat{\beta}\hat{\tau}}=\frac{2}{r^{2}}\left(2\delta^{\hat{\alpha}}_{\beta}\delta^{\alpha}_{\hat{\beta}}-\delta^{\hat{\alpha}}_{\hat{\beta}}\delta^{\alpha}_{\beta}\right)\delta^{\sigma}_{\tau}\delta^{\hat{\sigma}}_{\hat{\tau}}. \label{rslash}
\end{equation}
Armed with these results one may readily evaluate the portion of $\delta \chi^{mn}_{r}$ which depends on $R.$

Finally, we evaluate the term proportional to the field $D^{mn}_{rs}$ in the variation of $\chi^{mn}_{r}$ appearing in \eqref{dilatinov}. We make use of the ansatz \eqref{ansatz}-\eqref{so5sp4}, and find
\begin{equation}
D^{mn}_{rs}=D^{\hat{\alpha}\hat{\sigma}\hat{\beta}\hat{\tau}}_{\hat{\gamma}\hat{\upsilon}\hat{\zeta}\hat{\omega}}=\frac{d}{2}\left[\varepsilon^{\hat{\alpha}\hat{\beta}}\varepsilon_{\hat{\gamma}\hat{\zeta}}\left(\phantom{\int}\hspace{-.2in}3B^{\hat{\sigma}\hat{\tau}}B_{\hat{\upsilon}\hat{\omega}}-\delta^{\hat{\sigma}}_{\hat{\upsilon}}\delta^{\hat{\tau}}_{\hat{\omega}}-5\delta^{\hat{\sigma}}_{\hat{\omega}}\delta^{\hat{\tau}}_{\hat{\upsilon}}\right)+4\delta^{\hat{\alpha}}_{\hat{\zeta}}\delta^{\hat{\beta}}_{\hat{\gamma}}\left(\phantom{\int}\hspace{-.2in}\delta^{\hat{\sigma}}_{\hat{\upsilon}}\delta^{\hat{\tau}}_{\hat{\omega}} -\delta^{\hat{\sigma}}_{\hat{\omega}}\delta^{\hat{\tau}}_{\hat{\upsilon}}\right)\right]. \label{dexplicit}
\end{equation}
Comparing \eqref{rslash} and \eqref{dexplicit}, and applying the spinor constraint \eqref{spinorfactorized}, we find that $\delta \chi^{mn}_{r}$ vanishes provided that
\begin{equation}
d=\frac{3}{r^{2}}.
\end{equation}
This is the specialization of the general result stated in \eqref{constants} to the case of the round three-sphere.

\section{Zero Modes and the Three-Dimensional Action}
Our next task is to reduce the five-dimensional action to three dimensions.  The fields in the Yang-Mills multiplet consist of a gauge field $A,$ five scalars $\varphi_{\hat{A}}$ and a symplectic Majorana fermion $\rho^{m}.$   The action for these fields in the presence of supergravity background fields was derived in \cite{us}.  It takes the form of a sum of four terms
\begin{equation}
S=S_{A}+S_{\varphi}+S_{\rho}+S_{int},
\end{equation}
where each action is defined explicitly in \eqref{5dnonabelianaction}.

We proceed by expressing all five-dimensional fields as representations of the symmetry group $so(3)_{L}\times so(3)_{R}\times so(2)_{L}\times so(2)_{R}$.  Then, we identify zero-modes and carry out Kaluza-Klein reduction.  In section \ref{abelianzeromodes} we carry out the procedure for the for each of the various different fields in the five-dimensional Lagrangian.  In section \ref{rzero} we describe the subtle $r\rightarrow 0$ limit of the zero mode action and prove that the resulting low energy theory is indeed Chern-Simons theory with a complexified gauge group.
\subsection{Zero Modes}
\label{abelianzeromodes}
\subsubsection{Gauge Field}
The kinetic term for the five-dimensional gauge field takes the form of a standard Yang-Mills action with a Ramond-Ramond Chern-Simons coupling.
\begin{equation}
S_{A}=\frac{1}{8\pi^{2}}\int_{\mathbb{R}^{3}\times S^{2}}\left(\hspace{-.18in} \phantom{\int}\alpha \mathrm{Tr}(F\wedge * F)+G\wedge CS(A)\right),  \label{5dkingauge}
\end{equation}
where in the above $CS(A)$ is the Chern-Simons functional.

To carry out dimensional reduction, we note that the two-sphere admits no topologically non-trivial one-cycles.  Thus, in the reduction to three dimensions the only massless mode of $A$ which survives is the zero mode which is independent of position on the two-sphere.  We denote this three-dimensional field by $A_{b},$ where the index $b$ is now restricted to take the values $0,1,2$.  Its kinetic action is trivially obtained by integrating \eqref{5dkingauge} over the two-sphere.  The integral of $G$ over $S^{2}$ measures the Chern class of the Hopf fibration of $S^{3}$ and yields a Chern-Simons interaction.\footnote{In fact in our geometry the integral of $G$ over $S^{2}$ is $-2\pi$, leading to a level $-1$ interaction in 3d.  We change this to level 1 by a parity reflection on $A$.}
\begin{equation}
S_{A}=\frac{r \ell^{2}}{8\pi}\int_{\mathbb{R}^{3}} \mathrm{Tr}(F\wedge * F)+\frac{1}{4\pi}\int_{\mathbb{R}^{3}}CS(A) \label{SA3d}
\end{equation}
The most important feature of the above action is the very different scaling between the two terms.  The Yang-Mills interaction is suppressed by $r$, and hence in the dimensional reduction limit $r\rightarrow 0$ we can anticipate that the Chern-Simons term will dominate.

\subsubsection{Scalars}
Next we consider the kinetic action for scalar fields.  In five dimensions this takes the form
\begin{equation}
S_{\varphi}  = \frac{1}{32\pi^{2}} \int_{\mathbb{R}^{3}\times S^{2}} d^{5}x \sqrt{|g|} \ \alpha \mathrm{Tr}\left(\phantom{\int}\hspace{-.15in} \mathcal{D}_{a}\varphi^{mn}\mathcal{D}^{a}\varphi_{mn}-4\varphi^{mn}F_{ab}T^{ab}_{mn}-\varphi^{mn}(M_{\varphi})^{rs}_{mn} \varphi_{rs}\right). \label{s5dscalrs}
\end{equation}
As our background fields have simple properties in terms of the group $so(3)_{R}\times so(2)_{R}$ it is useful to convert the field $\varphi^{mn}$ from symplectic $sp(4)_{R}$ notation to $so(5)_{R}$ notation.  We write
\begin{equation}
\varphi^{mn}=\varphi_{\hat{A}}\left(\Gamma^{\hat{A}}\right)^{mn}
\end{equation}
The field $\varphi_{\hat{A}}$ is a vector in the $\mathbf{5}$ of $so(5)_{R}$.  As our background partially breaks the R symmetry, the effective action will distinguish between the various components of $\varphi_{\hat{A}}$.  Thus we write
\begin{equation}  
X_{\hat{a}}=\varphi_{\hat{a}},  \hspace{.5in}Y_{1}=\varphi_{3}, \hspace{.2in}Y_{2}=\varphi_{4}.
\end{equation}
The fields $Y_{1}, Y_{2}$ transform as a doublet under the unbroken $so(2)_{R}$ symmetry.  The fields $X_{\hat{a}}$ transform in the $\mathbf{3}$ of the unbroken diagonal subgroup of $so(3)_{R}\times so(3)_{L}$.  After reduction, this subgroup is identified with the Lorentz group in $\mathbb{R}^{3},$ and thus it is natural to think of $X_{\hat{a}}$ as comprising the components of a one-form.  We will see this explicitly in the action below.

We now seek to reduce to the action to one for the massless fields in $\mathbb{R}^{3}$.  In the action \eqref{s5dscalrs}, there are two sources of mass terms from the supergravity background fields.
\begin{itemize}
\item Explicit supergravity induces masses in the pairing $M_{\varphi}$ 
\begin{equation}
(M_{\varphi})^{rs}_{mn} =  \left[\left(\frac{1}{20\alpha^{2}}G_{ab}G^{ab}-\frac{R}{5}\right)\delta^{r}_{m}\delta^{s}_{n}+\frac{1}{2}\left(S^{r}_{[m}S^{s}_{n]}-S^{s}_{t}S^{t}_{[m}\delta^{r}_{n]}\right)-\frac{1}{15}D^{rs}_{mn}-T^{ab}_{mn}T^{rs}_{ab}\right]. \nonumber
\end{equation} 
\item Induced masses from the non-vanishing profile of the R gauge field.  Such terms occur in the in the square of the covariant derivative term in \eqref{s5dscalrs}.  They take the form
\begin{equation}
V_{a \ m }^{\phantom{a}[t}\delta_{n}^{u]}V_{a[t}^{\phantom{a} s}\delta^{r}_{u]} \nonumber.
\end{equation}
\end{itemize}
Upon summing these two contributions using our background field expressions appearing in \eqref{ansatz},  \eqref{so5sp4}, and \eqref{constants}, we find that the mass terms vanish identically for both the fields $X_{\hat{a}}$ and $Y_{z}$.

From the vanishing of the mass terms it follows that the structure of zero modes is very simple: the modes $X_{\hat{a}}$ and $Y_{z}$ which are constant on the sphere are massless.  All other scalar modes, associated to non-trivial profiles on $S^{2}$ have masses on the order of $1/r$ and hence decouple from the low-energy three-dimensional action in the limit $r\rightarrow 0$.

We may now calculate the effective action for the zero modes by simply integrating \eqref{s5dscalrs} over the two-sphere.  In terms of the fields $X$ and $Y$ the answer takes the form
\begin{eqnarray}
S_{X} & = & \frac{ r \ell^{2}}{8\pi}\int d^{3}x \mathrm{Tr}\left(\phantom{\int}\hspace{-.17in}\nabla_{a}X_{b}\nabla_{a}X_{b}\right)+\frac{1}{4\pi} \int d^{3}x \ i\varepsilon_{abc}\mathrm{Tr}\left(\phantom{\int}\hspace{-.17in}X_{a}\nabla_{b}X_{c}-i\sqrt{1-\ell^{2}}X_{a}F_{bc}\right), \nonumber\\
S_{Y} & = & \frac{ r \ell^{2}}{8\pi}\int d^{3}x \mathrm{Tr}\left(\phantom{\int}\hspace{-.17in}\nabla_{a}Y_{z}\nabla_{a}Y_{z}\right).\label{3dbosonaction}
\end{eqnarray}
In the above $\nabla_{a}$ indicates the ordinary gauge covariant derivative for adjoint valued fields
\begin{equation}
\nabla_{a}Z=\partial_{a}Z+[A_{a},Z],
\end{equation}
and we have removed the hat from the index of the field $X$ to emphasize the fact that in the unbroken diagonal subgroup $so(3)_{R}\times so(3)_{L}$ the field $X$ transforms as a one-form.

In \eqref{3dbosonaction} we again see terms with different scaling behaviors in the $r\rightarrow 0$ limit.  If one were to proceed naively, and simply set $r$ to zero in \eqref{3dbosonaction} one would conclude that the second order kinetic terms for $X$ and the entire action for $Y$ vanish.  We address this issue by carefully treating the $r\rightarrow 0$ limit in section \ref{rzero}.  
\subsubsection{Fermions}
Finally, we consider the most involved case of the fermion zero modes.  The five-dimensional action is
\begin{equation}
S_{\rho}  = \frac{1}{32\pi^{2}} \int d^{5}x  \sqrt{|g|}\ \alpha\mathrm{Tr}\left(\phantom{\int}\hspace{-.15in}  \rho_{m}i\slashed{\mathcal{D}}\rho^{m}+ \rho_{m}(M_{\rho})^{mn} \rho_{n} \right). \label{s5dferms}
\end{equation}
To begin, we make manifest the representation content of the field $\rho$ under $so(3)_{R}\times so(2)_{R}\times so(3)_{L}\times so(2)_{L}$.  Thus we express the field $\rho$ as
\begin{equation}
\rho^{m}=\varepsilon^{\alpha \hat{\alpha}}\lambda^{\sigma \hat{\sigma}}+\left(\gamma^{a}\right)^{\alpha \hat{\alpha}}\xi_{a}^{\sigma \hat{\sigma}}.
\end{equation}
The decomposition above is completely general and involves no restriction on $\rho^{m}$.  The variables $\lambda$ and $\xi_{a}$ are independent five-dimensional Grassmann fields.  As their index structure indicates, from the point of view of the unbroken diagonal subgroup of $so(3)_{R}\times so(3)_{L}$, the field $\lambda$ transforms as a scalar, while the field $\xi_{a}$ transforms as a one-form. 

We now seek to reduce the action to one for the massless fermions in $\mathbb{R}^{3}$.  In the action \eqref{s5dferms} there are three sources of mass terms that we must take into account.
\begin{itemize}
\item Explicit supergravity induced mass terms in the pairing $M_{\rho}$
\begin{equation}
(M_{\rho})^{mn} = \left[\frac{1}{2}S^{mn}+\frac{1}{8\alpha}\slashed{G} \ \Omega^{mn}-\frac{1}{2}\slashed{T}^{mn}\right]. \nonumber
\end{equation}
\item Induced masses from the non-vanishing profile of the R gauge field.  Such terms arise from the covariant derivative and take the form
\begin{equation}
-\frac{i}{2}\slashed{V}^{m}_{n}. \nonumber
\end{equation}
\item Curvature induced masses from the non-trivial spin connection on the two-sphere.
\end{itemize}

To properly take account of the curvature induced mass terms we decompose the fermions according to their profile in the sphere.  Let $\slashed{D}_{S^{2}}$ dentate the Dirac operator on the two-sphere, and let $\kappa$ indicate the two-dimensional chirality matrix.  A convenient basis of modes are spinors $\Theta_{j}^{\sigma}$ which are eigenfunctions of the operator $\kappa \slashed{D}_{S^{2}},$ 
\begin{equation}
\kappa^{\sigma}_{\upsilon}\left(\slashed{D}_{S^{2}}\right)^{\upsilon}_{\tau}\Theta^{\tau}=\nu\Theta^{\sigma},
\end{equation}
where in the above, $\nu$ is the eigenvalue of the spinor $\Theta$.  The chirality matrix $\kappa$ anti-commutes with the Dirac operator and squares to the identity hence 
\begin{equation}
\left(\kappa \slashed{D}_{S^{2}}\right)\left(\kappa \slashed{D}_{S^{2}}\right)=-(\kappa \kappa )\left(\slashed{D}_{S^{2}}\slashed{D}_{S^{2}}\right)=-\left(\slashed{D}_{S^{2}}\right)^{2}.
\end{equation}
It follows that the spectrum of $\kappa \slashed{D}_{S^{2}}$ is related to that of the Dirac operator by multiplication by $i$.  Thus the eigenvalues $\nu$ take the form
\begin{equation}
\nu=\frac{2n}{r\ell}, \hspace{.5in}n \in \mathbb{Z}, \hspace{.5in}n\neq0.
\end{equation}
The modes $\Theta^{\sigma}$ satisfy an orthonormality condition.  We have
\begin{equation}
\int_{S^{2}}d^{2}x\sqrt{g}\ \Theta^{\sigma}\widetilde{\Theta}^{\tau}B_{\sigma \tau}\propto \delta(\nu+\widetilde{\nu}), \hspace{.5in}\int_{S^{2}}d^{2}x\sqrt{g} \ \Theta^{\sigma} \widetilde{\Theta}^{\tau}\kappa_{\sigma \tau}\propto \delta(\nu-\widetilde{\nu}),
\end{equation}

We may now sum the contributions to the three-dimensional fermion masses.  We find that there are massless three-dimensional fields contained in the sector with eigenvalues $\nu=\pm2/r\ell$.  Modes with $|\nu|$ larger than this minimal value are massive with masses of order $1/r$ and decouple from the low-energy effective action.  

Our next task is to isolate the effective action for the massless fields.  Each relevant eigenvalue $\nu=\pm 2/r\ell$ is degenerate with multiplicity two.  We define spinors  $a_{i}^{\sigma}$ and $b_{i}^{\sigma}$ to label the associated modes
\begin{equation}
\kappa^{\sigma}_{\upsilon}\left(\slashed{D}_{S^{2}}\right)^{\upsilon}_{\tau}a_{i}^{\tau}=\left(\frac{2}{r\ell}\right)a_{i}^{\sigma}, \hspace{.5in}\kappa^{\sigma}_{\upsilon}\left(\slashed{D}_{S^{2}}\right)^{\upsilon}_{\tau}b_{i}^{\tau}=-\left(\frac{2}{r\ell}\right)b_{i}^{\sigma}.
\end{equation}
The index $i=\pm$ labels distinct solutions to the above equations.  We may choose this index to label the behavior of solutions under a rotation of the angle $\phi$
\begin{equation}
J_{\phi}a_{\pm}^{\sigma}= \pm \frac{i}{2}a_{\pm}^{\sigma}, \hspace{.5in}J_{\phi}b_{\pm}^{\sigma}= \pm \frac{i}{2}b_{\pm}^{\sigma}.
\end{equation}
Explicit mode functions are 
\begin{eqnarray}
a_{+}^{\sigma}=\frac{e^{i\phi/2}}{\sqrt{4\pi r\ell^{2} }}\left[\begin{array}{c}i\sin(\theta/2) \\ \cos(\theta/2)\end{array}\right], \hspace{.5in}a_{-}^{\sigma}=\frac{e^{-i\phi/2}}{\sqrt{4\pi r\ell^{2}}}\left[\begin{array}{c}\cos(\theta/2) \\ i\sin(\theta/2)\end{array}\right], \\
b_{+}^{\sigma}=\frac{e^{i\phi/2}}{\sqrt{4\pi r\ell^{2} }}\left[\begin{array}{c}-i\sin(\theta/2) \\ \cos(\theta/2)\end{array}\right], \hspace{.5in}b_{-}^{\sigma}=\frac{e^{-i\phi/2}}{\sqrt{4\pi r\ell^{2}}}\left[\begin{array}{c}\cos(\theta/2) \\ -i\sin(\theta/2)\end{array}\right]. \nonumber
\end{eqnarray}
The non-vanishing pairings between these modes are given by
\begin{eqnarray}
\int_{S^{2}}d^{2}x\sqrt{g} \ a_{i}^{\sigma}b_{j}^{\tau}B_{\sigma \tau} & = &\left(\frac{r}{4}\right)B_{ij}, \\ 
\int_{S^{2}}d^{2}x\sqrt{g}\ a_{i}^{\sigma}a_{j}^{\tau}\kappa_{\sigma \tau}=\int_{S^{2}}d^{2}x\sqrt{g} \ b_{i}^{\sigma}b_{j}^{\tau}\kappa_{\sigma \tau} & = &-\left(\frac{r}{4}\right)\varepsilon_{ij}. \nonumber
\end{eqnarray}

With these preliminaries we may now carry out the truncation of the five-dimensional action to a theory of a finite number of light three-dimensional fields.  We take an ansatz of the form
\begin{eqnarray}
\xi_{a}^{\sigma \hat{\sigma}} & = & \xi_{a}^{i \hat{i}}\delta_{\hat{i}}^{\hat{\sigma}}a_{i}^{\sigma}+\widetilde{\xi}_{a}^{i \hat{i}}\delta_{\hat{i}}^{\hat{\sigma}}b_{i}^{\sigma}, \label{fermionzeros} \\
\lambda^{\sigma \hat{\sigma}} & = & \lambda^{i \hat{i}}\left(a_{i}^{\sigma}\delta_{\hat{i}}^{\hat{\sigma}}-eb_{j}^{\sigma}\kappa^{j}_{i}\kappa_{\hat{i}}^{\hat{\sigma}}\right)+ \widetilde{\lambda}^{i \hat{i}}\left(b_{i}^{\sigma}\delta_{\hat{i}}^{\hat{\sigma}}+ea_{j}^{\sigma}\kappa^{j}_{i}\kappa_{\hat{i}}^{\hat{\sigma}}\right). \nonumber
\end{eqnarray}
In the above, the quantities $\lambda^{i\hat{i}}, \widetilde{\lambda}^{i\hat{i}},$  $\xi_{a}^{i\hat{i}},$ and $\widetilde{\xi}_{a}^{i\hat{i}}$ are the independent three dimensional fermionic fields, and the coefficients in the expansions appearing in \eqref{fermionzeros} are chosen to diagonalize the mass matrix.   Each of the indices $i, \hat{i}$ appearing on these fields takes on two possible values.  Thus, from the point of view of the unbroken diagonal subgroup of $so(3)_{R}\times so(3)_{L},$ we have identified eight Grassmann valued scalars: the $\lambda$ and $\widetilde{\lambda}$ fields, as well as eight Grassmann valued one-forms, the $\xi$ and $\widetilde{\xi}$ fields.

Upon substitution of \eqref{fermionzeros} into the action \eqref{s5dferms} we find that the $\lambda$ modes are massless, while the other fields $\widetilde{\lambda},$ $\xi,$ and $\widetilde{\xi}$ are massive with masses of order $1/r$.   Naively, one might expect that we may neglect these heavy fermions.  However, dropping such fields is not justified due to the fact that the heavy fields pair directly with the massless $\lambda$ field at the level of the kinetic term.  Indeed, we find that the action takes the form
\begin{eqnarray}
S_{ferm} & = & \frac{1}{32\pi^{2}} \int d^{3}x \  \mathrm{Tr}\left[\left(\phantom{\int}\hspace{-.16in}\xi_{a}^{i\hat{i}}\varepsilon_{ij}B_{\hat{i}\hat{j}}-e\widetilde{\xi}_{a}^{i\hat{i}}B_{ij}\varepsilon_{\hat{i}\hat{j}}\right)i\nabla_{a}\lambda^{j\hat{j}}\right. \label{sferm3d}\\
& - & \left.\frac{i}{r\ell}\left(\phantom{\int}\hspace{-.16in}\xi_{a}^{i\hat{i}}\xi_{a}^{j\hat{j}}-\widetilde{\xi}_{a}^{i\hat{i}}\widetilde{\xi}_{a}^{j\hat{j}}\right)\varepsilon_{ij}B_{\hat{i}\hat{j}} -\frac{4i}{r\ell^{2}(1+\ell)}\widetilde{\lambda}^{i\hat{i}}\widetilde{\lambda}^{j\hat{j}}\varepsilon_{ij}B_{\hat{i}\hat{j}}\right]. \nonumber
\end{eqnarray}
The action \eqref{sferm3d} has a number of features which merit comment.
\begin{itemize}
\item The fields $\widetilde{\lambda},$ $\xi,$ and $\widetilde{\xi}$ are indeed massive with masses of order $1/r$.  The field $\lambda$ is massless.
\item The kinetic term of the massless field $\lambda$ arises from a pairing with massive $\xi$ and $\widetilde{\xi}$ fields.  Thus, one cannot simply discard the massive modes and set them to zero.  Similarly, we will see that the field $\widetilde{\lambda}$ mixes with the massless mode $\lambda$ via the Yukawa coupling with the massless field $Y$, and hence $\widetilde{\lambda}$ cannot simply be discarded.   
\item In arriving at \eqref{sferm3d} we have dropped the kinetic terms for the massive fields.  It is legitimate to discard these terms because in the $r \rightarrow 0$ limit the kinetic terms are dominated by the mass terms.
\end{itemize}
We address the $r\rightarrow 0$ limit of the action in detail in section \ref{rzero}.

\subsubsection{Non-Abelian Interaction Terms}
\label{nonabcouplings}
The final piece of the action contains non-abelian interaction terms.  In five dimensions it takes the form
\begin{equation}
S_{int} = \frac{1}{32\pi^{2}} \int d^{5}x  \sqrt{|g|}\ \alpha \mathrm{Tr}\left(\phantom{\int}\hspace{-.15in}\rho_{m\alpha}[\varphi^{mn},\rho_{n}^{\alpha}]-\frac{1}{4}[\varphi_{mn},\varphi^{nr}][\varphi_{rs},\varphi^{sm}]-\frac{2}{3}S_{mn} \varphi^{mr}[\varphi^{ns},\varphi_{rs}]\right) \label{s5dintred1}
\end{equation}

Given our identification of zero modes in the previous sections it is straightforward to reduce the above terms to three dimensions by integrating over the sphere.  The scalar potential terms reduce to 
\begin{eqnarray}
S_{pot} &= &\frac{r \ell^{2}}{8\pi}\int d^{3}x \  \mathrm{Tr}\left(\frac{1}{2}[X_{a},X_{b}][X_{a},X_{b}]+[X_{a},Y_{z}][X_{a},Y_{z}]+\frac{1}{2}[Y_{z},Y_{w}][Y_{z},Y_{w}]\right)  \nonumber \\
& + & \frac{i \sqrt{1-\ell^{2}}}{12\pi} \int d^{3}x \ i\varepsilon_{abc}\mathrm{Tr}\left(\phantom{\int}\hspace{-.18in}X_{a}[X_{b},X_{c}]\right) \label{s3dpot}
\end{eqnarray}

We similarly reduce the Yukawa couplings.  Retaining only those terms which couple directly to the massless fermion $\lambda$ we find
\begin{equation}
S_{yuk} = \frac{1}{32\pi^{2}}\int d^{3}x  \ \mathrm{Tr}\left(\phantom{\int}\hspace{-.16in}\widetilde{\xi}_{a}^{i\hat{i}}[X_{a},\lambda^{j\hat{j}}]B_{ij}\varepsilon_{\hat{i}\hat{j}}-e\xi_{a}^{i\hat{i}}[X_{a},\lambda^{j\hat{j}}]\varepsilon_{ij}B_{\hat{i}\hat{j}}+\left(\frac{2}{1+\ell}\right)\widetilde{\lambda}^{i\hat{i}}[Y_{z},\lambda^{j\hat{j}}]B_{ij}\kappa^{z}_{\hat{i}\hat{j}}\right). \label{s3dyuk}
\end{equation}
\subsection{The $3d$ Effective Action: Complex Chern-Simons Theory}
\label{rzero}
Let us take stock of our results thus far.  The three-dimensional effective action enjoys a symmetry under $so(3)\times so(2)_{R}$.  The $so(3)$ is interpreted as the three-dimensional rotation symmetry although its five-dimensional origin is the as the diagonal subgroup of $so(3)_{R}\times so(3)_{L}$.  The $so(2)_{R}$ symmetry is inherited from the R symmetry of the 5d Yang-Mills theory.  With respect to these symmetries the field content of the model consists of the following.
\begin{itemize}
\item Bosons.   

There is a gauge field $A$ for a $\mathfrak{g}$ gauge symmetry.  There are three scalars $X$ transforming as a one-form under $so(3)$ and as singlets under $so(2)_{R}$.  There are two scalars $Y$ transforming as singlets under $so(3)$ and as a doublet under $so(2)_{R}.$
\item Fermions.

There are four massless fermion fields $\lambda^{i\hat{i}}$ transforming as scalars under $so(3)$ and as a pair of doublets under $so(2)_{R}$.
\end{itemize}
The effective action of these fields is stated in equations \eqref{SA3d}, \eqref{3dbosonaction}, \eqref{sferm3d}, \eqref{s3dpot}, and \eqref{s3dyuk}.  We now complete our analysis of the $r\rightarrow0$ limit.  

\subsubsection{The Fermion Action}
As our first step we consider the portion of the action which involves the fermions $\lambda$.  This takes the form
\begin{eqnarray}
S_{ferm} +S_{yuk}& = & \frac{1}{32\pi^{2}} \int d^{3}x \  \mathrm{Tr}\left[\left(\phantom{\int}\hspace{-.16in}\xi_{a}^{i\hat{i}}\varepsilon_{ij}B_{\hat{i}\hat{j}}-e\widetilde{\xi}_{a}^{i\hat{i}}B_{ij}\varepsilon_{\hat{i}\hat{j}}\right)i\nabla_{a}\lambda^{j\hat{j}}\right. \label{stotlatfermmass}\\
& - & \left.\frac{i}{r\ell}\left(\phantom{\int}\hspace{-.16in}\xi_{a}^{i\hat{i}}\xi_{a}^{j\hat{j}}-\widetilde{\xi}_{a}^{i\hat{i}}\widetilde{\xi}_{a}^{j\hat{j}}\right)\varepsilon_{ij}B_{\hat{i}\hat{j}} -\frac{4i}{r\ell^{2}(1+\ell)}\widetilde{\lambda}^{i\hat{i}}\widetilde{\lambda}^{j\hat{j}}\varepsilon_{ij}B_{\hat{i}\hat{j}}\right. \nonumber \\
& + & \left.\widetilde{\xi}_{a}^{i\hat{i}}[X_{a},\lambda^{j\hat{j}}]B_{ij}\varepsilon_{\hat{i}\hat{j}}-e\xi_{a}^{i\hat{i}}[X_{a},\lambda^{j\hat{j}}]\varepsilon_{ij}B_{\hat{i}\hat{j}} +\left(\frac{2}{1+\ell}\right)\widetilde{\lambda}^{i\hat{i}}[Y_{z},\lambda^{j\hat{j}}]B_{ij}\kappa^{z}_{\hat{i}\hat{j}}\right].\nonumber
\end{eqnarray}

The fields $\xi$, $\widetilde{\xi}$ and $\widetilde{\lambda}$ are massive with masses that tend to infinity as $r\rightarrow0$.  However, one may not simply set such fields to zero due to their quadratic coupling to the massless field $\lambda$.  Instead, we must integrate out the heavy fermions exactly.  This can readily be done in the $r\rightarrow0$ limit because in that limit the presence of the parametrically large mass terms means that quantum fluctuations are suppressed and we may simply solve the equations of motion for these fields.  Such equations of motion take the form
\begin{eqnarray}
\xi_{a}^{j\hat{j}}& = &\frac{r\ell}{2}\nabla_{a}\lambda^{j\hat{j}} -\frac{ir\ell e}{2}[\lambda^{j\hat{j}},X_{a}], \nonumber \\
\widetilde{\xi}_{a}^{j\hat{j}} & = & \frac{r\ell e}{2}\nabla_{a}\lambda^{k\hat{k}}\kappa_{k}^{j}\kappa_{\hat{k}}^{\hat{j}}-\frac{ir\ell}{2}[\lambda^{k\hat{k}},X_{a}]\kappa_{k}^{j}\kappa_{\hat{k}}^{\hat{j}}, \\
\widetilde{\lambda}^{j\hat{j}} & = & \frac{ir\ell^{2}}{4}[Y_{z},\lambda^{k\hat{k}}]\kappa_{k}^{j}\kappa^{z\hat{j}}_{\phantom{z}\hat{k}}. \nonumber
\end{eqnarray}
Upon substituting into the action \eqref{stotlatfermmass} and simplifying one obtains the following effective action for the massless fields $\lambda$
\begin{eqnarray}
S_{\lambda} & = &\frac{ir\ell^{2}}{64\pi^{2}(1+\ell)}\int d^{3}x \ \mathrm{Tr}\left(\phantom{\int}\hspace{-.17in}\nabla_{a}\lambda^{i\hat{i}}\nabla_{a}\lambda^{j\hat{j}}\varepsilon_{ij}B_{\hat{i}\hat{j}}+[X_{a},\lambda^{i\hat{i}}][X_{a},\lambda^{j\hat{j}}]\varepsilon_{ij}B_{\hat{i}\hat{j}}\right. \label{Slambda} \\
& - & \left. \frac{1}{2}[Y_{z},\lambda^{i\hat{i}}] [Y_{w},\lambda^{j\hat{j}}] \left(\delta^{zw}\varepsilon_{ij}B_{\hat{i}\hat{j}}+i\varepsilon^{zw}\varepsilon_{ij}\varepsilon_{\hat{i}\hat{j}}\right)\phantom{\int}\hspace{-.2in}\right). \nonumber
\end{eqnarray}
Note that the $\ell$ dependence of the fermion action has become trivial.  It can be absorbed by a dimensionless rescaling of the field $\lambda$.

Although \eqref{Slambda} has been obtained in a straightforward way, it contains a striking feature: it is second order in derivatives for the Grassmann valued scalars $\lambda$.  One context where such actions naturally arise is as the Faddeev-Popov action for ghosts which gauge fix a gauge redundancy.  As we will argue in section \ref{ghosts}, this is indeed the correct interpretation of the fields $\lambda$.

\subsubsection{Supersymmetry Transformations and a $Q$ Exact Deformation}
To proceed further in our analysis it is necessary to study the supersymmetry transformations of our model.  Such transformations are inherited from the five-dimensional variations applied to the zero modes.  They take the simple form
 \begin{eqnarray}
\delta A_{b} & = & \beta^{i\hat{i}}\varepsilon_{ij}B_{\hat{i}\hat{j}}\nabla_{b}\lambda^{j\hat{j}}, \nonumber \\
\delta X_{b} & = & \beta^{i\hat{i}}\varepsilon_{ij}B_{\hat{i}\hat{j}}[X_{b},\lambda^{j\hat{j}}] ,\\
\delta Y_{z} & = & \beta^{i\hat{i}}[Y_{w},\lambda^{j\hat{j}}]\left(\delta_{zw}\varepsilon_{ij}B_{\hat{i}\hat{j}}+i\varepsilon_{zw}\varepsilon_{ij}\varepsilon_{\hat{i}\hat{j}}\right), \nonumber \\
\delta \lambda ^{i\hat{i}}& = & 8\pi i(1+\ell)[Y_{z},Y_{w}]\varepsilon_{zw}\kappa^{\hat{i}}_{\hat{j}}\beta^{i\hat{j}}, \nonumber
\end{eqnarray}
where in the above $\beta^{i\hat{i}}$ is a Grassmann coefficient of the transformation.

Let us focus our attention on the variation of the gauge field $A_{b}$ and the scalars $X_{b}$.  Acting on these fields, the supersymmetry transformations behave as gauge transformations with gauge parameter  
\begin{equation}
\beta^{i\hat{i}}\varepsilon_{ij}B_{\hat{i}\hat{j}}\lambda^{j\hat{j}}.
\end{equation}
In particular, this means that supersymmetric observables constructed from only $X$ and $A$ are simply the gauge invariant functions of these fields.  From now on, we restrict ourselves to the study of such gauge invariant, i.e. supersymmetric observables which do not depend in any way on the variables $\lambda$ and $Y$.  Our aim is therefore to integrate out these fields.

At first sight this task may seem hopeless.  The bosonic action involves non-linear functions of the scalars, for example quartic interactions $\mathrm{Tr}\left([Y_{z},Y_{w}][Y_{z},Y_{w}]\right)$, as well as nonlinear couplings between the fermions $\lambda$ and the scalars $Y$.  However, supersymmetry allows us to overcome this difficulty.   Again the key observation is that the supersymmetry variation acts non-trivially only on the $\lambda$ and $Y$ fields.  As a result the offending non-linear terms are an exact supersymmetry variation.  Specifically 
\begin{equation}
\delta^{i\hat{i}}\left(\Xi_{i\hat{i}}\right) = [Y_{z},\lambda^{i\hat{i}}] [Y_{w},\lambda^{j\hat{j}}] \left(\delta^{zw}\varepsilon_{ij}B_{\hat{i}\hat{j}}+i\varepsilon^{zw}\varepsilon_{ij}\varepsilon_{\hat{i}\hat{j}}\right)+8\pi i(1+\ell)[Y_{z},Y_{w}][Y_{z},Y_{w}], \label{nonlinear}
\end{equation}
where $\Xi_{i\hat{i}}$ is given by
\begin{equation}
\Xi_{i\hat{i}} \sim \mathrm{Tr}\left(\phantom{\int}\hspace{-.17in}\lambda_{i\hat{j}} \kappa^{\hat{j}}_{\hat{i}} [Y_{z},Y_{w}]\varepsilon_{zw}\right).
\end{equation}

This result has the following useful implication.  If we are interested in supersymmetric, i.e. gauge invariant observables of $X$ and $A$ alone then the presence of the nonlinear terms  \eqref{nonlinear} does not effect their resulting expectation values.  Hence, we may freely delete these terms from the action.  The action for the bosons $Y$ and the fermions $\lambda$ is then quadratic, and therefore the exact functional integral over these fields can be performed.  We carry out this integration in section \ref{ghosts}

\subsubsection{Ghosts and Gauge Fixing}
\label{ghosts}
We are now ready to complete our derivation and take the $r\rightarrow 0$ limit.  As we have previously emphasized, we are interested only in the gauge invariant observables which are functions of $A$ and $X$ alone.  Thus we consider the action for the fields $Y$ and $\lambda$.  Up to an overall dimensionless coefficient it is given by
\begin{equation}
S_{ghost}=r\int d^{3}x \ \mathrm{Tr}\left(\phantom{\int}\hspace{-.17in}\nabla_{a}\lambda^{i\hat{i}}\nabla_{a}\lambda^{j\hat{j}}\varepsilon_{ij}B_{\hat{i}\hat{j}}+\nabla_{a}Y_{z}\nabla_{a}Y_{z}+[X_{a},\lambda^{i\hat{i}}][X_{a},\lambda^{j\hat{j}}]\varepsilon_{ij}B_{\hat{i}\hat{j}}+[X_{a},Y_{z}][X_{a},Y_{z}]\right).
\end{equation}
The action $S_{ghost}$ vanishes in the $r\rightarrow 0$ limit, however as it is quadratic in the field variables, the exact path integral can be done and the $r\rightarrow 0$ effects understood.  

We note that there are four fermions $\lambda^{i\hat{i}}$ and a pair of real scalars $Y_{z}$, and that the action for the $\lambda$'s is identical to the action for the scalars.  It follows that the functional integral over the $Y$ variables cancels the integral over two of the $\lambda$ fields.  The net result is then a simple functional determinant.
\begin{equation}
\int \mathcal{D}Y \mathcal{D}\lambda \ e^{-S_{ghost}[\lambda,Y]}=\mathrm{det}\left(\phantom{\int}\hspace{-.16in}\nabla_{a}^{2}+\left(\mathrm{ad}_{X_{a}}\right)^{2}\right), \label{fpdet}
\end{equation}
where in the above $\mathrm{ad}_{X_{a}}$ denotes the operator defined by $X_{a}$ acting on fields in the adjoint representation.

Having successfully integrated out the fields $Y$ and $\lambda$ we obtain an action for the fields $A$ and $X$ supplemented by the determinant \eqref{fpdet} in the functional integral measure.  Thus our field theory is defined by the path integral
\begin{equation}
\int \mathcal{D}X \mathcal{D}A  \ \mathrm{det}\left(\phantom{\int}\hspace{-.16in}\nabla_{a}^{2}+\left(\mathrm{ad}_{X_{a}}\right)^{2}\right) e^{-S[X,A]}, \label{measure}
\end{equation}
where the action $S[X,A]$ takes the form of a sum of two terms, one of which is $r$ independent and one of which vanishes as $r\rightarrow 0$.  The $r$ independent piece is 
\begin{equation}
 \frac{1}{4\pi}\int_{\mathbb{R}^{3}}\left(CS(A)+ i\varepsilon_{abc}\mathrm{Tr}\left(\phantom{\int}\hspace{-.17in}X_{a}\nabla_{b}X_{c}\right)\right)
-\frac{\sqrt{\ell^{2}-1}}{4\pi} \int d^{3}x \ i\varepsilon_{abc}\mathrm{Tr}\left(\phantom{\int}\hspace{-.18in}X_{a}F_{bc}-\frac{2}{3}X_{a}X_{b}X_{c}\right). \label{csfirst}
\end{equation}
While the piece which vanishes as $r\rightarrow 0$ can be written as
\begin{equation}
 r \int d^{3}x \  \mathrm{Tr}\left(F\wedge * F+\nabla_{a}X_{b}\nabla_{a}X_{b}+\frac{1}{2}[X_{a},X_{b}][X_{a},X_{b}]\right). \label{complexym}
\end{equation}

The action \eqref{csfirst}-\eqref{complexym}, together with the modified path integral measure \eqref{measure} constitutes our final answer for the resulting 3d quantum field theory.  To complete our analysis, we now claim that it is possible to interpret this result as a complex Chern-Simons theory with gauge group $\mathfrak{g}_{\mathbb{C}}$ together with a specific choice of contour of integration in field space.  The $r$ dependent terms in the action may then be understood as a specific regulator of the complex Chern-Simons theory.

To illustrate these claims, we first simplify the action using the following manipulations.

\begin{itemize}
\item First, in the $r$ dependent action \eqref{complexym} make use of the identity
\begin{equation}
\int d^{3}x  \ \mathrm{Tr}\left(\phantom{\int}\hspace{-.16in}\nabla_{a}X_{b}\nabla_{a}X_{b}\right)=\int d^{3}x  \ \mathrm{Tr}\left(\phantom{\int}\hspace{-.16in}2\nabla_{[a}X_{b]}\nabla_{[a}X_{b]}-X_{a}X_{b}F_{ab}+\nabla_{a}X_{a}\nabla_{b}X_{b}\right).
\end{equation}

\item  Next, redefine the field $X_{a}\rightarrow iX_{a}.$   This redefinition may interpreted as a contour prescription in the resulting complex Chern-Simons theory.
\end{itemize}

After these steps the entire action takes a simple elegant form.  We introduce a complexified $\mathfrak{g}_{\mathbb{C}}$ gauge field
\begin{equation}
\mathcal{A}_{a}=A_{a}+iX_{a},
\end{equation}
and denote by $\mathcal{F}_{ab}$ the associated complex field strength.  Then the $r$ independent action takes the form
\begin{equation}
 \frac{q}{8\pi} \int \mathrm{Tr}\left({\cal A} \wedge d {\cal A} + \frac{2}{3} {\cal A} \wedge {\cal A} \wedge {\cal A}\right) + \frac{\tilde{q}}{8\pi} \int \mathrm{Tr}\left(\bar{\cal A} \wedge d\bar{\cal A} + \frac{2}{3} \bar{\cal A} \wedge \bar{\cal A} \wedge\bar{\cal A} \right),\end{equation}
where the levels are
\begin{equation}
q=1+i\sqrt{1-\ell^{2}}, \hspace{.5in}\tilde{q}=1-i\sqrt{1-\ell^{2}}. \label{finallevels}
\end{equation}
Meanwhile the $r$ dependent piece of the action \eqref{complexym} also has an simple expression.  It takes the form of the real part of the complex Yang-Mills term plus a familiar correction
\begin{equation}
\mathrm{Re}\left[r\int_{\mathbb{R}^{3}}d^{3}x \ \mathrm{Tr}\left(\phantom{\int}\hspace{-.17in}\mathcal{F}_{ab}\mathcal{F}_{ab}\right)\right] -r \int_{\mathbb{R}^{3}}d^{3}x\ \mathrm{Tr}\left(\phantom{\int}\hspace{-.17in}(\nabla_{a}X_{a})^{2}\right). \label{gaugefixed}
\end{equation}

Observe that, with the exception of the term depending on the divergence $\nabla _{a}X^{a}$, the entire action is invariant under complexified $\mathfrak{g}_{\mathbb{C}}$ gauge transformations.  It is therefore natural to interpret the divergence term in \eqref{gaugefixed} as a gauge fixing term.  Under the non-compact part of the $\mathfrak{g}_{\mathbb{C}}$ gauge transformations the fields $X$ and $A$ transform as
\begin{equation}
\delta X_{a}= \nabla_{a}g, \hspace{.5in} \delta A_{a}=[g,X_{a}].
\end{equation}
Under these transformations, the variation of the candidate gauge fixing term $\nabla_{a}X_{a}$ is
\begin{equation}
\delta\left( \nabla_{a}X_{a}\right)=\nabla^{2}g+\left(\mathrm{ad}_{X}\right)^{2}g.
\end{equation}
It follows that the Faddeev Popov determinant for this gauge fixing term is exactly the modified measure we obtained in \eqref{measure}.  Thus, to understand the $r\rightarrow0$ limit, we may simply undo the gauge fixing and drop the complex Yang-Mills terms.  In this way we obtain complex Chern-Simons theory at the levels stated in \eqref{finallevels}.
\section{Discussion}
\label{discussion}

We have shown that the supersymmetric reduction of the $(2,0)$ on
the squashed three-sphere is $\mathfrak{g}_{\mathbb{C}}$ Chern-Simons theory. This
establishes the equality of the $S^3_\ell$ partition function of
the 3d ${\cal N}=2$ SCFT $T_\frak{g}(M_3)$ with the Chern-Simons
partition function on general three-manifolds $M_3.$ The importance of supersymmetry lay in the
independence of the 6d $S^3_\ell \times M_3$ partition function on
the ratio of sizes, enabling us to reduce the calculation to one in 5d Yang-Mills.

The limit $\ell \rightarrow \infty$ of a squashed sphere with very
small fiber results in very weakly coupled 5d Yang-Mills. The 3d
Chern-Simons level, $\sqrt{1-\ell^2} \rightarrow \infty$, and that
theory becomes weakly coupled. Thus we predict that the logarithm
$S^3_\ell$ partition function of $T_{\frak{su}(2)}(M_3)$ in that
limit reproduces the volume of the 3-manifold, which is the
classical limit of $SL(2,\mathbb{C})$ Chern-Simons theory.

The partition function of 3d ${\cal N}=2$ quantum field theories
 on the ellipsoid geometry, $S^3_b$, equals that of the squashed sphere with $\ell =
\frac{2}{b+b^{-1}}$ \cite{Imamura:2011wg}. Note that this only covers the range $\ell
\leq 1$ of the squashing parameter.  It was expected that the $b \rightarrow 0$ limit of the conjectured effective
3d Chern-Simons theory would be weakly coupled \cite{Terashima:2011xe}, with level of
order $\frac{1}{b^2}$. However, in \eqref{finallevels} this limit results in
levels 1 and $i$.

It seems possible that there is a type of s-duality that relates
this to weak coupling \cite{Dimofte:2011jd}. In particular, the squashed sphere with
$\ell < 1$ has a large fiber, so it is natural to reduce on a
different contractible circle in the base $S^2$. Doing so, one can
follow the same procedure as before, and find the 5d
supersymmetric background, now with varying dilaton. In this flipped reduction, the Yang-Mills coupling will shrink as
$\ell \rightarrow 0$. One might conjecture that the reduction to
3d will again result in $\mathfrak{g}_{\mathbb{C}}$ Chern-Simons, but with an
s-dual value of the levels.

Our results may also shed light on the AGT correspondence \cite{Alday:2009aq, Alday:2009fs}.  That conjecture relates the $S^4$ partition function \cite{Pestun:2007rz} of the 4d ${\cal N}=2$ theory $T_N[\Sigma_g]$ of $N$ M5 branes on a Riemann surface $\Sigma_g$ to the partition function on $\Sigma_g$ of the rank $N$ Toda theory. The geometry of $S^4$ can be thought of as an $S^3$ fibered over an interval, $I$, and shrinking at the two ends. This decomposition is compatible with the preserved supersymmetries of the ${\cal N}=2$ theory on $S^4$. 

Therefore roughly speaking, one may reduce first on the $S^3$ to obtain $SL(N,\mathbb{C})$ Chern-Simons on $\Sigma_g \times I$ with appropriate boundary conditions.  As shown in \cite{Henneaux:2010xg} the resulting field theory on $\Sigma_g$ is precisely the rank $N$ Toda theory, generalizing the fact that Liouiville theory appears on the boundary of $SL(2,\mathbb{C})$ Chern-Simons, since the latter is equivalent to 3d gravity \cite{Witten:1988hc, Coussaert:1995zp}. 
  
\section*{Acknowledgements}

We thank C. Vafa, and X. Yin for discussions.  The work of C.C. is support by a Junior Fellowship at the Harvard Society of Fellows.  The work of D.J. is supported by the Fundamental Laws Initiative Fund at Harvard University, and the National Science Foundation Grant No. 1066293.  

\appendix
\section{5d Vector Multiplets in Supergravity Backgrounds}
\label{5dsymdetails}
In this section we state the relationship between the 6d background supergravity fields and the 5d supergravity fields.   In addition, we state the action and supersymmetry transformations for 5d maximally supersymmetric Yang-Mills theory in off-shell supergravity backgrounds. These results were obtained in \cite{us}, and they are summarized here for completeness.  

The 6d supergravity theory \cite{Bergshoeff:1999db} is invariant under a Weyl rescaling symmetry as well as an $so(5)$ R symmetry.  The scaling dimension of a field will be indicated by $w$.  To facilitate calculations with spinors we view the R symmetry group equivalently as $sp(4)$, and let indices $m,n ,r \cdots $ range form 1 to 4 and indicate objects in the fundamental $\mathbf{4}$ of $sp(4)$. The antisymmetric second rank invariant tensor of $sp(4)$ is denoted as $\Omega_{mn}.$ It may be used to raise and lower symplectic indices
\begin{equation}
\Xi^{n}=\Omega^{nm}\Xi_{m}, \hspace{.5in}\Xi_{n}=\Xi^{m}\Omega_{mn}, \hspace{.5in}\Omega_{mn}=-\Omega_{nm}.
\end{equation}
The 6d supergravity multiplet contains the bosonic fields indicated in Table \ref{6dgfields}.
\begin{table}[h]
\centering
\begin{tabular}{|c|c|c|c|c|}
\hline
Field & Interpretation & Restriction & $sp(4)$ & w \\
\hline
\multirow{2}{*}{$\underline{e}_{\mu}^{a}$}  & \multirow{2}{*}{Metric } & \multirow{2}{*}{coframe}& \multirow{2}{*}{$\mathbf{1}$} & \multirow{2}{*}{-1} \\
& & & & \\
\hline
\multirow{2}{*}{$\underline{V}^{mn}_{\mu}$} & \multirow{2}{*}{R Gauge Field} & \multirow{2}{*}{$\underline{V}^{mn}_{\mu}=\underline{V}^{nm}_{\mu}$} & \multirow{2}{*}{$\mathbf{10}$} &\multirow{2}{*}{ 0} \\
& & & & \\
\hline
\multirow{2}{*}{$ \underline{T}^{mn}_{\mu \nu \rho}$}& \multirow{2}{*}{Auxiliary 3-form} & $\underline{T}^{mn}=-*\underline{T}^{mn}$ & \multirow{2}{*}{$\mathbf{5}$} &\multirow{2}{*}{ -2 }\\
 & &$\underline{T}^{mn}=-\phantom{\frac{1}{1_{a_{a_{a}}}}}\hspace{-.27in}\underline{T}^{nm}, \hspace{.2in}\phantom{A^{A^{A^{A}}}}\Omega_{mn}\underline{T}^{mn}=0.$ & & \\
\hline
\multirow{2}{*}{$ \underline{D}^{mn,rs}$} &\multirow{2}{*}{ Auxiliary scalar} & $\underline{D}^{mn,rs}=\underline{D}^{rs,mn}=-\underline{D}^{nm,rs}=-\underline{D}^{mn,sr},$ & \multirow{2}{*}{$\mathbf{14}$} & \multirow{2}{*}{2}\\
 & &$\Omega_{mn}\underline{D}^{mn,rs}=\Omega_{rs}\underline{D}^{mn,rs}=\Omega_{mr}\Omega_{ns}\underline{D}^{mn,rs}=0.$ & & \\
 \hline
\end{tabular}
\caption{Bosonic fields of 6d (2,0) off shell supergravity}
\label{6dgfields}
\end{table}

Upon dimensional reduction the 6d mertric degrees of freedom are reduced to a 5d metric, graviphoton, and dilaton in the standard fashion
\begin{equation}
\underline{e}^{a}_{\mu}=\left(\begin{array}{cc}e^{a}_{\mu} & e^{5}_{\mu}=\alpha^{-1}C_{\mu} \\
e^{a}_{z}=0 & e^{5}_{z}=\alpha^{-1}
\end{array} \right), \label{coframered}
\end{equation}
 while the remaining 6d bosons descend as  
\begin{eqnarray}
\underline{V}_{a}^{mn} & \rightarrow &\begin{cases} V_{a}^{mn} & a \neq 5  \\
\underline{V}_{5}^{mn}\equiv S^{mn} & \end{cases}, \nonumber\\
\underline{T}^{mn}_{abc} & \rightarrow & \underline{T}^{mn}_{ab5} \equiv T^{mn}_{ab},\\
\underline{D}^{mn,rs} & \rightarrow & D^{mn,rs}. \nonumber
\end{eqnarray}
These account for the 5d bosonic background fields indicated in Table \ref{5dgravfields}.

The conditions for a 5d background to preserve supersymmetry are that the variations of the fermions in the supergravity multiplet vanish.  These variations take the form
\begin{eqnarray}
\delta \psi_{a}^{m} & = & \mathcal{D}_{a}\epsilon^{m}+\frac{i}{2\alpha}\left[\phantom{\frac{1}{1}}\hspace{-.11in}G_{ab}\Omega^{mn}-\alpha S^{mn}\eta_{ab}\right]\Gamma^{b}\epsilon_{n}+\frac{i}{8\alpha}\left[\phantom{\frac{1}{1}}\hspace{-.11in}G^{bc}\Omega^{mn}-4\alpha\left(T^{mn}\right)^{bc}\right]\Gamma_{abc}\epsilon_{n}, \nonumber \\
\delta \chi^{mn}_{r} & = &\left[\phantom{\frac{1}{1}}\hspace{-.1in}T^{mn}_{ab}T_{cdrs}-\frac{1}{\alpha}T^{mn}_{ab}G_{cd}\Omega_{rs}+\frac{1}{12}\left(\mathcal{D}^{e}S^{[m}_{r}\delta^{n]}_{s}+\mathcal{D}_{f}T^{mn f e}\Omega_{rs}\right)\varepsilon_{eabcd}\right]\Gamma^{abcd}\epsilon^{s} \nonumber \\
& + &\left[ \frac{5}{2\alpha}T^{mn}_{ab}G^{a}_{\phantom{a}c}\Omega_{rs}-4 T^{mn}_{ab}T^{a}_{\phantom{a}crs}+2T^{mn}_{bc}S_{rs}- S_{p}^{[m}T^{n]p}_{bc}\Omega_{rs}-R_{bc r}^{\phantom{b}[m}\delta^{n]}_{s}\right.  \label{5dsusyvar} \\
& + & \left.\frac{1}{2}\mathcal{D}_{a}T^{mn}_{de}\Omega_{rs}\varepsilon^{ade}_{\phantom{ade}bc}\right]\Gamma^{bc}\epsilon^{s}+\left[\frac{1}{\alpha}T^{mn}_{ab}G^{ab}\Omega_{rs}-2 T^{mn}_{ab}T^{ab}_{rs}-\frac{4}{15}D^{mn}_{rs}\right]\epsilon^{s}  - (\mathrm{traces}), \nonumber
\end{eqnarray}
where in the above the notation ``traces'' indicates terms proportional to $sp(4)$ invariant tensors $\Omega_{mn}$ and $\delta_{n}^{m}$, and the five-dimensional covariant derivatives, curvatures, and connections are
\begin{eqnarray}
\mathcal{D}_{\mu}\epsilon^{m} & = & \partial_{\mu}\epsilon^{m}+\frac{1}{2}\partial_{\mu}\log(\alpha)\epsilon^{m}+\frac{1}{4}\omega_{\mu}^{bc}\Gamma_{bc}\epsilon^{m}-\frac{1}{2}V_{\mu n}^{m}\epsilon^{n},  \nonumber\\
\mathcal{D}_{\mu}S^{m n} & = & \partial _{\mu}S^{mn}-\partial_{\mu}\log(\alpha)S^{mn}-V_{\mu r}^{(m}S^{n)r}, \nonumber \\
\mathcal{D}_{\mu}T^{mn}_{ab} & = & \partial_{\mu}T^{mn}_{ab}-2\omega_{\mu [a}^{c}T^{mn}_{b]c}-\partial_{\mu}\log(\alpha)T^{mn}_{ab} + V_{\mu s}^{[m}T^{n]s}_{ab}, \label{5dcurvdefs}\\
R_{\mu \nu}^{mn} & = & 2\partial_{[\mu}V^{mn}_{\nu]}+V_{[\mu}^{r(m}V_{\nu]r}^{n)}, \nonumber \\
\omega_{\mu}^{ab} & = &  2e^{\nu[a}\partial_{[\mu}e_{\nu]}^{\hspace{.1in}b]} -e^{\rho[a}e^{b]\sigma}e_{\mu}^{c}\partial_{\rho}e_{\sigma c}+2e_{\mu}^{[a}\partial^{b]}\log(\alpha). \nonumber
\end{eqnarray}

The action for the vector multiplet in the supergravity background is a sum of the following four terms.
\begin{eqnarray}
S_{A} & = &\frac{1}{8\pi^{2}}\int \mathrm{Tr} \left(\phantom{\int}\hspace{-.15in}\alpha F\wedge *F+C\wedge F \wedge F\right), \nonumber\\
S_{\varphi} & = &\frac{1}{32\pi^{2}} \int d^{5}x \sqrt{|g|} \ \alpha \mathrm{Tr}\left(\phantom{\int}\hspace{-.15in} \mathcal{D}_{a}\varphi^{mn}\mathcal{D}^{a}\varphi_{mn}-4\varphi^{mn}F_{ab}T^{ab}_{mn}-\varphi^{mn}(M_{\varphi})^{rs}_{mn} \varphi_{rs}\right),  \nonumber\\
S_{\rho} & = &\frac{1}{32\pi^{2}} \int d^{5}x  \sqrt{|g|}\ \alpha\mathrm{Tr}\left(\phantom{\int}\hspace{-.15in}  \rho_{m\gamma}i\slashed{\mathcal{D}}^{\gamma}_{\beta}\rho^{m\beta}+ \rho_{m\gamma}(M_{\rho})^{mn \gamma}_{\phantom{mn}\beta} \rho^{\beta}_{n} \right). \label{5dnonabelianaction}\\
S_{int} & = &\frac{1}{32\pi^{2}} \int d^{5}x  \sqrt{|g|}\ \alpha \mathrm{Tr}\left(\phantom{\int}\hspace{-.15in}\rho_{m\alpha}[\varphi^{mn},\rho_{n}^{\alpha}]-\frac{1}{4}[\varphi_{mn},\varphi^{nr}][\varphi_{rs},\varphi^{sm}]-\frac{2}{3}S_{mn} \varphi^{mr}[\varphi^{ns},\varphi_{rs}]\right).\nonumber
\end{eqnarray}
In the above the covariant derivatives acting on $\rho$ and $\varphi$ are defined as
\begin{eqnarray}
\mathcal{D}_{\mu}\rho^{m} & = & \left(\partial_{\mu}-\frac{3}{2}\partial_{\mu}\log(\alpha)+\frac{1}{4}\omega^{bc}_{\mu}\Gamma_{bc}\right)\rho^{m}-\frac{1}{2}V_{\mu n}^{m}\rho^{n}+[A_{\mu},\rho^{m}],\label{nonabeliancovardmat} \\
\mathcal{D}_{\mu}\varphi_{mn} & = & \left(\phantom{\int}\hspace{-.15in}\partial_{\mu}-\partial_{\mu}\log(\alpha)\right)\varphi_{mn}-V_{\mu[m}^{\phantom{\mu}r}\varphi_{n]r}+[A_{\mu},\varphi_{mn}].\nonumber \\
 \nonumber
\end{eqnarray}
And the supergravity induced mass terms are
\begin{eqnarray}
(M_{\varphi})^{rs}_{mn} & = &  \left[\left(\frac{1}{20\alpha^{2}}G_{ab}G^{ab}-\frac{R}{5}\right)\delta^{r}_{m}\delta^{s}_{n}+\frac{1}{2}\left(S^{r}_{[m}S^{s}_{n]}-S^{s}_{t}S^{t}_{[m}\delta^{r}_{n]}\right)-\frac{1}{15}D^{rs}_{mn}-T^{ab}_{mn}T^{rs}_{ab}\right] , \nonumber\\
(M_{\rho})^{mn \alpha}_{\phantom{mn}\beta}  & = &\left[\frac{1}{2}S^{mn}\delta^{\alpha}_{\beta}+\frac{1}{8\alpha}G_{ab}\left(\Gamma^{ab}\right)^{\alpha}_{\beta}\Omega^{mn}-\frac{1}{2}T^{mn}_{ab}\left(\Gamma^{ab}\right)^{\alpha}_{\beta}\right].  \label{sugramass}
\end{eqnarray}
Finally, the supersymmetry variations of the vector multiplet fields are
\begin{eqnarray}
\delta A_{c} & = &- \frac{i}{4}\epsilon_{m}\Gamma_{c}\rho^{m}, \nonumber \\
\delta \varphi^{mn} & = &- \epsilon^{[m}\rho^{n]}-\frac{1}{4}\Omega^{mn}\epsilon^{r}\rho_{r}, \label{5dsynonmat}\\
\delta \rho^{m} & = & \left(\phantom{\int}\hspace{-.15in}S^{[m}_{s}\varphi^{n]s}\Omega_{rn}-2\varphi^{mn}S_{nr}-i\slashed{\mathcal{D}}\varphi^{mn}\Omega_{rn}\right)\epsilon^{r} -\frac{1}{2}\Omega_{nr}[\varphi^{mn}, \varphi^{rs}]\epsilon_{s}\nonumber \\
& + & \frac{1}{4}\left(\phantom{\int}\hspace{-.15in} 2F^{ab}\delta^{m}_{r}-\varphi^{ns}T_{ns}^{ab}\delta^{m}_{r}-4\varphi^{mn}T_{nr}^{ab}-\frac{2}{\alpha}\varphi^{mn}G^{ab}\Omega_{rn}\right)\Gamma_{ab}\epsilon^{r}.\nonumber
\end{eqnarray}

\section{Clifford Algebra}
\label{cliffordcovnetions}
\subsection{Reducing $5d$ Spinors to $3d\otimes 2d$ Spinors}
\label{5=3+2}
To analyze the spherical backgrounds studied in this paper it is necessary to reduce the spinor of $so(5)$ to a representation of the group $so(3)\times so(2)$.  This is achieved as follows.  The spinor $\mathbf{4}$ of $so(5)$ reduces to a product $(\mathbf{2},\mathbf{2})$ of Dirac spinors of $so(3)\times so(2)$.  Thus we may reduce the five-dimensional Dirac algebra to a tensor product form.
\begin{equation}
\Gamma^{0}=\gamma^{0}\otimes\kappa, \hspace{.5in}\Gamma^{1}=\gamma^{1}\otimes\kappa, \hspace{.5in}\Gamma^{2}=\gamma^{2}\otimes\kappa, \hspace{.5in}\Gamma^{3}=\mathbf{1}_{2}\otimes\kappa^{3}, \hspace{.5in}\Gamma^{4}=\mathbf{1}_{2}\otimes\kappa^{4}.
\end{equation}
Here, the matrices $\gamma^{0}, \gamma^{1}, \gamma^{2}$ form a three-dimensional Clifford algebra of Lorentz signature, while the matrices $\kappa^{3}, \kappa^{4}$ constitute a two-dimensional Clifford algebra of Euclidean signature.  Finally, $\kappa$ is the two dimensional chirality matrix with eigenvalues $\pm1$. As explicit matrices we choose
\begin{equation}
\kappa^{3}=\left(\begin{array}{cc} 0 & -i \\  i & 0\end{array}\right), \hspace{.5in}\kappa^{4}=\left(\begin{array}{cc} 0 & 1 \\  1 & 0\end{array}\right), \hspace{.5in}\kappa=\left(\begin{array}{cc} 1 & 0 \\  0 & -1\end{array}\right). \label{2dcliffdefs}
\end{equation}

The five-dimensional charge conjugation matrix $C$ matrix is used to relate the spinor of $so(5)$ to its dual as
\begin{equation}
\Psi^{\alpha}=C^{\alpha \beta}\Psi_{\beta}, \hspace{.5in}\Psi_{\alpha}=\Psi^{\beta}C_{\beta \alpha}.
\end{equation}
Upon reduction to $so(3)\times so(2)$, the charge conjugation matrix reduces to
\begin{equation}
C=\left(\begin{array}{cc} 0 & 1 \\
-1 & 0
\end{array}
\right)\otimes \left(\begin{array}{cc}  0 & 1 \\ 1 & 0\end{array}\right).
\end{equation}
The above is the simply the tensor product of the invariant tensors $\varepsilon$ and $B$ of $so(3)$ spinors and $so(2)$ spinors which are discussed in section \ref{32facts}.

\subsection{Useful Properties of $3d$ and $2d$ Clifford Algebra}
\label{32facts}
In three or two Euclidean dimensions the Clifford algebra obeys a number of useful identities which we catalog here.
\subsubsection{$3d$}
The spinor of $so(3)$ is a two-dimensional representation with spinors indicated by $\lambda^{\alpha}$ where $\alpha =1,2$.  As a representation of $so(3)$ the spinor representation is isomorphic to its dual.  Hence there exists an invariant tensor of $so(3)$ which relates spinors with raised and lowered indices.  This invariant is the antisymmetric symbol $\varepsilon_{\alpha \beta}$.  Our conventions for the spinor $\varepsilon$ symbol are
\begin{equation}
\lambda^{\alpha}=\varepsilon^{\alpha \beta}\lambda_{\beta},\hspace{.4in}\lambda_{\alpha}=\lambda^{\beta}\varepsilon_{ \beta \alpha}, \hspace{.5in} \varepsilon_{12}=1.
\end{equation}
The gamma matrices act on spinors as $\lambda^{\alpha}\rightarrow(\gamma^{a}) ^{\alpha}_{\beta}\lambda^{\beta}$.  The matrices $(\gamma^{a}) ^{\alpha}_{\beta}$ are Hermitian while those with all lowered or raised indices, $(\gamma^{a})^{\alpha \beta}$ or $(\gamma^{a})_{\alpha \beta}$ are symmetric.

Finally there are several contraction identities and rearrangement formulas
\begin{eqnarray}
(\gamma^{a})^{\alpha}_{\sigma}(\gamma^{b})^{\sigma}_{\beta} & =  &\delta^{ab}\delta^{\alpha}_{\beta}+i\varepsilon^{abc}(\gamma_{c})^{\alpha}_{\beta},  \nonumber\\
(\gamma^{a})^{\alpha}_{\beta}(\gamma^{b})^{\sigma}_{\tau} & =  & \varepsilon^{\alpha \sigma}(\gamma^{a}\gamma^{b})_{\beta \tau}+(\gamma^{a})^{\sigma}_{\beta}(\gamma^{b})^{\alpha}_{\tau},\\
(\gamma^{a})^{\alpha}_{\beta}(\gamma_{a})^{\sigma}_{\tau} & = & 2 \delta^{\alpha}_{\tau}\delta^{\sigma}_{\beta}-\delta^{\alpha}_{\beta}\delta^{\sigma}_{\tau}. \nonumber
\end{eqnarray}
\subsubsection{$2d$}
The Dirac spinor of $so(2)$ is two-dimensional representation with spinors indicated by $\lambda^{\sigma}$ where $\sigma =+,-$, labels Weyl subspaces of definite chirality.  As a representation of $so(2)$ the Dirac spinor representation is isomorphic to its dual.  Hence there exists an invariant tensor of $so(2)$ which relates spinors with raised and lowered indices.  This invariant is the symmetric symbol $B_{\sigma \tau}$.  Our conventions for the spinor $B$ symbol are given by
\begin{equation}
\lambda^{\sigma}=B^{\sigma \tau}\lambda_{\tau},\hspace{.4in}\lambda_{\sigma}=\lambda^{\tau}B_{ \tau \sigma}, \hspace{.5in} B_{12}=B_{21}=1, \hspace{.5in} B_{11}=B_{22}=0.
\end{equation}
The Clifford algebra matrices act on spinors as $\lambda^{\sigma}\rightarrow(\kappa^{a}) ^{\sigma}_{\tau}\lambda^{\tau}$.  The matrices $(\kappa^{a}) ^{\sigma}_{\tau}$ are Hermitian while those with all lowered or raised indices, $(\kappa^{a})^{\sigma \tau}$ or $(\kappa^{a})_{\sigma \tau}$ are symmetric.

The chirality matrix $\kappa$ anti-commutes with all $\kappa^{a}$ and has eigenvalues $\pm 1$ on the positive and negative chirality subspace respectively.  Upon raising or lowering an index, the chirality matrix becomes equivalent to the antisymmetric symbol
\begin{equation}
B^{\upsilon \tau}\kappa^{\sigma}_{\tau}=\kappa^{\upsilon\sigma }=-\varepsilon^{\upsilon\sigma }, \hspace{.5in}\kappa^{\sigma}_{\tau}B_{ \sigma\upsilon}=\kappa_{ \tau\upsilon}=\varepsilon_{ \tau\upsilon}.
\end{equation}
Finally, we have the following contraction identity
\begin{equation}
(\kappa^{a})^{\sigma}_{\chi}(\kappa_{a})^{\psi}_{\upsilon}=B^{\sigma \psi}B_{\chi\upsilon}+\delta^{\sigma}_{\upsilon}\delta^{\psi}_{\chi}-\delta^{\sigma}_{\chi}\delta^{\psi}_{\upsilon}.
\end{equation}

\bibliography{Arxiv.bbl}{}
\bibliographystyle{utphys}

\end{document}